\documentclass[fleqn,usenatbib]{mnras}

\usepackage{newtxtext,newtxmath}

\usepackage[T1]{fontenc}
\usepackage{ae,aecompl}

\usepackage{graphicx}	
\usepackage{amsmath}	
\usepackage{amssymb}	
\def\ha{H$\alpha$}
\def\hb{H$\beta$}
\def\oiii{[O\,{\sc iii}]}
\def\oi{[O\,{\sc i}]}
\def\oii{[O\,{\sc ii}]}
\def\sii{[S\,{\sc ii}]}
\def\siii{[S\,{\sc iii}]}
\def\nii{[N\,{\sc ii}]}
\def\kms{$\rm km\,s^{-1}$}

\def\arc{$^{\prime\prime}$}
\def\h2{H$_2$}
\def\feii{[Fe\,{\sc ii}]}
\def\feiip{[Fe\,{\sc ii}]}
\def\feiii{[Fe\,{\sc iii}]}
\def\fevii{[Fe\,{\sc vii}]}
\def\fevi{[Fe\,{\sc vi}]}

\def\heii{He\,{\sc ii}}

\def\hei{He\,{\sc i}}
\def\pii{[P\,{\sc ii}]}
\def\brg{Br${\sc \gamma}$}
\def\pab{Pa${\sc \beta}$}
\def\ariii{[Ar\,{\sc iii}]}
\def\arv{[Ar\,{\sc v}]}
\def\ariv{[Ar\,{\sc iv}]}
\def\cav{[Ca\,{\sc v}]}
\def\cliii{[Cl\,{\sc iii}]}

\title[IFS of NGC\,4395]{A close look at the dwarf AGN of NGC\,4395: optical and near-IR integral field spectroscopy}

\author[Brum et al.]{Carine Brum,$^{1}$\thanks{E-mail:
carinefisica@gmail.com} Marlon R. Diniz,$^{1}$  Rogemar A. Riffel,$^{1}$  Alberto Rodr\'iguez-Ardila,$^{2,3}$ 
\newauthor  Luis C. Ho,$^{4,5}$ Rog\'erio Riffel,$^6$ Rachel Mason$^{7,8}$, Lucimara Martins,$^9$ 
Andreea Petric,$^{10,7}$ 
\newauthor Rub\'en S\'anchez-Janssen$^{11}$ \\
$^{1}$Universidade Federal de Santa Maria, Departamento de F\'\i sica, CCNE, 97105-900, Santa Maria, RS, Brazil\\
$^2$ Laborat\'orio Nacional de Astrof\'isica - Rua dos Estados Unidos 154, Bairro das Na\c c\~oes. CEP 37504-364, Itajub\'a, MG, Brazil\\
$^{3}$ Instituto de Astrof\'isica de Canarias, C/V\'ia L\'actea s/n, E-38205, La Laguna, Tenerife, Spain.\\
$^4$ Kavli Institute for Astronomy and Astrophysics, Peking University, Beijing 100871, People's Republic of China\\
$^5$Department of Astronomy, School of Physics, Peking University, Beijing 100871, People's Republic of China\\
$^{6}$ Universidade Federal do Rio Grande do Sul, Instituto de F\'\i sica, CP 15051, Porto Alegre 91501-970, RS, Brazil\\
$^7$Gemini Observatory, Northern Operations Center, 670 North A'ohoku Place, Hilo, HI 96720, USA\\
$^8$Department of Plant \& Soil Science, University of Vermont, Burlington VT 05405, USA \\
$^9$NAT-Universidade Cruzeiro do Sul, Rua Galv\~ao Bueno, 868, 01506-000 S\~ao Paulo, SP, Brazil\\
$^{10}$ Canada-France-Hawaii Telescope, 65-1238 Mamalahoa Highway, Kamuela, HI, 96743, USA\\
$^{11}$ STFC UK Astronomy Technology Centre, Royal Observatory, Blackford Hill, Edinburgh, EH9 3HJ, UK.
} 

\begin{document}

\pagerange{\pageref{firstpage}--\pageref{lastpage}} \pubyear{2016}

\maketitle

\label{firstpage}

\begin{abstract})
Intermediate mass black holes (10$^3$--10$^5$\,M$_\odot$) in the center of dwarf galaxies are believed to be analogous to growing Active Galactic Nuclei (AGN) in the early Universe. Their characterization can provide insight about the early galaxies. We present optical and near-infrared integral field spectroscopy of the inner $\sim$50~pc of the dwarf galaxy NGC\,4395, known to harbor an AGN. NGC\,4395 is an ideal candidate to investigate the nature of dwarf AGN, as it is nearby ($d\approx4.4$~Mpc) enough to allow a close look at its nucleus. The optical data were obtained with the Gemini GMOS-IFU covering the 4500\,\AA~ to 7300\,\AA\ spectral range at a spatial resolution of 10 pc. The J and K-band spectra were obtained with the Gemini NIFS at spatial resolutions of $\sim$5~pc. The gas kinematics show a compact, rotation disk component with a projected velocity amplitude of 25\,\kms. We estimate a mass of $7.7\times10^5$\,M$_\odot$ inside a radius of 10~pc. From the H$\alpha$ broad line component, we estimate the AGN bolometric luminosity as $L_{\rm bol}=(9.9\pm1.4)\times\,10^{40}$~erg\,s$^{-1}$ and a mass $M_{\rm BH}=(2.5^{+1.0}_{-0.8})\times10^5$\,M$_\odot$ for the central black hole.  The mean surface mass densities for the ionized and molecular gas are in the ranges (1--2)\,M$_{\odot}\,$pc$^{-2}$ and (1--4)$\times10^{-3}$\,M${_\odot}$\,pc$^{-2}$ and the average ratio between ionized and hot molecular gas masses is $\sim$500. The emission-line flux distributions reveal an elongated structure at 24\,pc west of the nucleus, which is blueshifted relative to the systemic velocity of the galaxy by $\approx$30\,\kms. We speculate that this structure is originated by the accretion of a gas-rich small satellite or by a low metallicity cosmic cloud.

\end{abstract}


\section{Introduction}

One of the most notable astrophysical discoveries of the past 3 decades is that supermassive black
holes (SMBHs) with masses of $M_{\rm BH}\sim10^{7}-10^{9}$ M$_{\odot}$  are common -- maybe ubiquitous -- 
at the center of massive galaxies \citep[e.g.][]{kh13}. Certain fundamental questions about SMBHs, however, remain to be answered: 
 How  were the `seeds'' of black holes formed? 
 How have SMBHs grown over cosmic time?  Have they influenced the evolution of their host galaxies, and if so, how?
Massive galaxies are thought to have grown through a succession of mergers with other galaxies,
and efficient accretion of large amounts of merger-supplied gas accounts for the bulk of the black
hole growth in the Universe. These processes, however, rapidly obscure the  formation history these SMBHs. They may not represent the initial growth of a SMBH from its 
$10^{3}-10^{5}$\,M$_\odot$ seed, erasing the effects of a small nascent Active Galactic Nucleus (AGN)
on its surroundings. 

Simulations suggest that radiative feedback from primordial ``miniquasars" shapes the star formation around
them and strongly suppresses their subsequent growth \citep{jeon}. Such processes are
impossible to study in high-redshift objects with current facilities. AGN in local dwarf galaxies,
though, offer an opportunity to study relatively small SMBHs ($\sim$10$^{5}$-10$^{6}$\,M$_{\odot}$)
that may be growing in a way that is analogous to the first  kind of such objects. 
Despite this utility, the first significant numbers of AGN-hosting dwarf galaxies have only 
recently been discovered \citep{greene04,greene07,dong,reines13}, as the community has concentrated its efforts 
on massive galaxies and the galaxy-SMBH co-evolution suspected on the basis of the well-known M$_{BH}$-$\sigma$
relation \citep[e.g.][]{gebhardt00,ferrarese00,ferrarese05,kh13}. However the association of intermediate black holes with AGN is still in debate and a recent work by \citet{chilingarian18} confirmed the AGN nature of only ten sources of a sample of 305 candidate galaxies of harboring intermediate mass black hole. For example, high quality integral field spectroscopy of He\,2-10, a well known candidate of harboring an intermediate mass black hole, show no sign of an central AGN \citep{cresci17}.

Near-infrared (near-IR) and optical Integral Field Unit (IFU) observations have been carried out on numerous Seyfert AGN at resolutions of a few $\times$10 pc, the spatial scales thought necessary to probe fueling and feedback in low-luminosity objects \citep{hicks}. This has uncovered, for example inflows of molecular \citep[e.g.][]{n4051,ms09,ms18a,mrk79,davies14} and ionized \citep[e.g.][]{allan14b} gas and outflows of ionized gas  sometimes much larger than the AGN accretion rate 
\citep[e.g.][]{mrk79,baaa,lena,ms11,ms18b,ardila17}.  Hints of a relation between starburst age and Eddington ratio have also been observed, which suggest that star formation activity predates AGN activation by $\approx$100 Myr, and that stellar winds have a influence on AGN fueling \citep{davies}. Rings of intermediate-age stars, consistent with a scenario in which the formation of the stellar ring triggered an episode of nuclear activity are seen for some galaxies \citep{mrk1066-pop,rogerio11,diniz17}, evidence of relatively young nuclear stellar populations and nuclear gas reservoirs in Seyferts that are not detected in quiescent galaxies \citep{hicks,mallmann18}.

As a first step towards understanding the growth of small SMBHs, detailed analysis of the gas kinematics ionization structure and gas morphology must be carried out in dwarfs hosting AGNs.
In this context, NGC\,4395 is a dwarf spiral galaxy located at a distance $d\approx4.4$~Mpc (1$^{\prime\prime}\approx$20 pc) \citep{brok}, whose broad optical emission lines, X-ray variability and radio jet unambiguously reveal the
presence of an accreting SMBH \citep{fili89, shih, wrobel}. \citet{reppeto} using SDSS imaging estimated its stellar mass as 3.82$\times10^8$\,M$_{\odot}$. A striking characteristic of NGC\,4395 is that it is a bulgeless spiral galaxy, harboring a Seyfert 1 nucleus, which defies the correlation between the mass of the central black hole and the mass of the bulge that suggests that the growth of the black hole and the galaxy bulge are coupled. In addition NGC~4395 is one of the lowest mass galaxies with a well determined dynamical mass for its central black hole \citep{brok}. This makes NGC\,4395 unique and worthy of a detailed study of its central region.  

Reverberation mapping sets the mass of the SMBH of NGC\,4395 as $M_{\rm BH} (3.6\pm1.1)\,\times\,10^{5}$\,M$_{\odot}$ \citep{peterson}. 
A similar value of $M_{\rm BH}$ is also obtained by modeling the dynamics of the molecular gas in NGC\,4395, resulting in 
  $M_{\rm BH}=\,4{_{-3}^{+8}}\,\times\,10^{5}$\,M$_{\odot}$ \citep{brok}.
An optical continuum image obtained with the Hubble Space Telescope (HST) using the Wide-Field Planetary Camera 2 (WFPC2) and F606W filter reveals an emission structure (hereafter ``blob") at 1\farcs2 of the nucleus \citep{martini}, 
which has been interpreted as an ionized gas cloud whose line emission lies within the bandwidth of the F606W filter 
\citep{brok}. However, the nature of this blob is still not clear and its study represents a secondary motivation to obtain high-resolution integral field spectroscopy (IFS) of the central region of NGC\,4395, which allows not only the mapping the gas emission, but also its kinematics. 


This work is organized as follows: Section 2 presents a description of the observations and data reduction 
procedures. Sec. 3 shows the emission-line flux distributions, line-ratio maps, velocity fields and velocity 
dispersion maps. These maps are interpreted and discussed in  Sec. 4 and the conclusions of this work are presented in Sec. 5.
 
\begin{figure*}
\begin{tabular}{c c}
\includegraphics[width=0.43\textwidth]{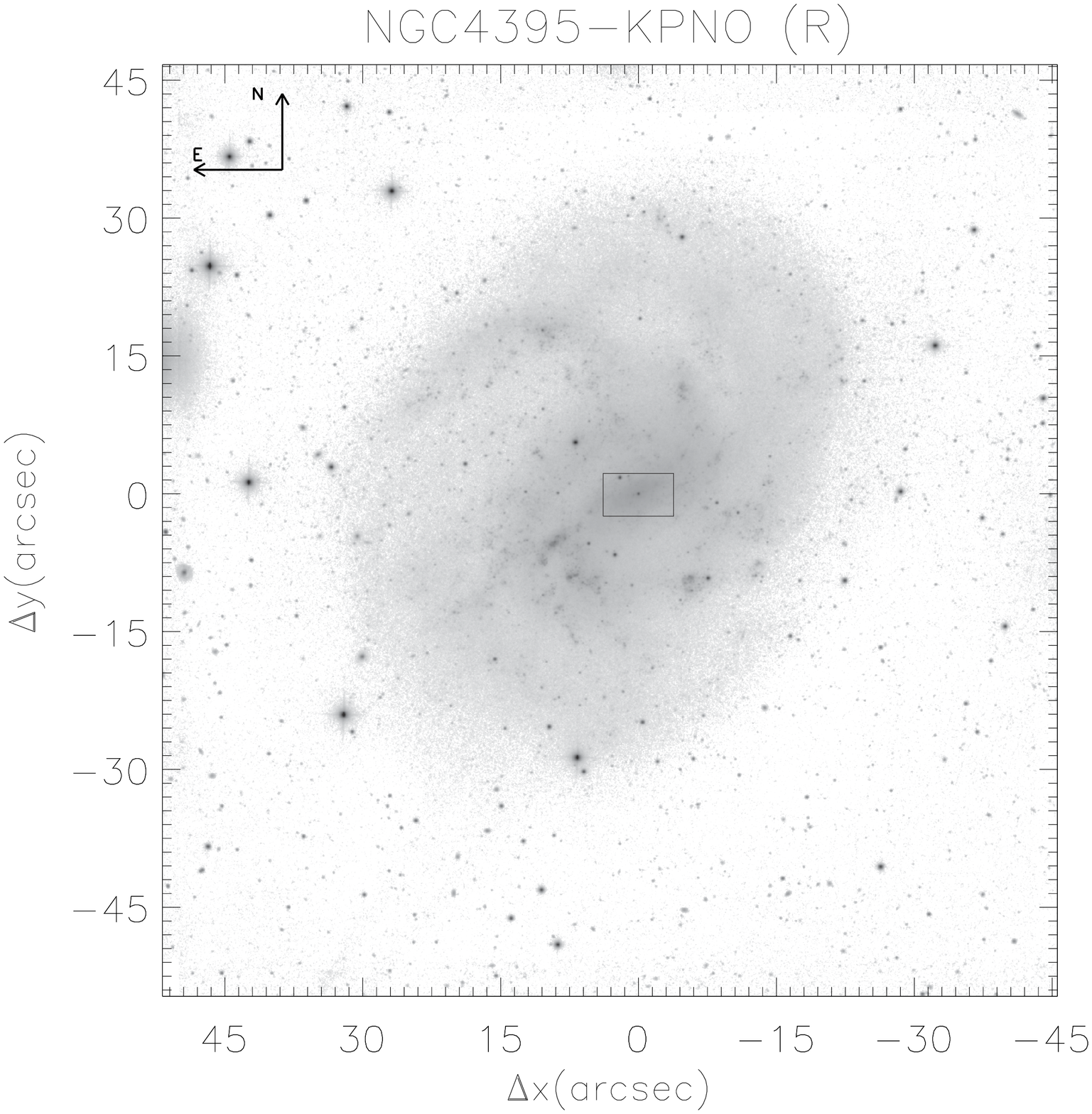} &
\includegraphics[width=0.43\textwidth]{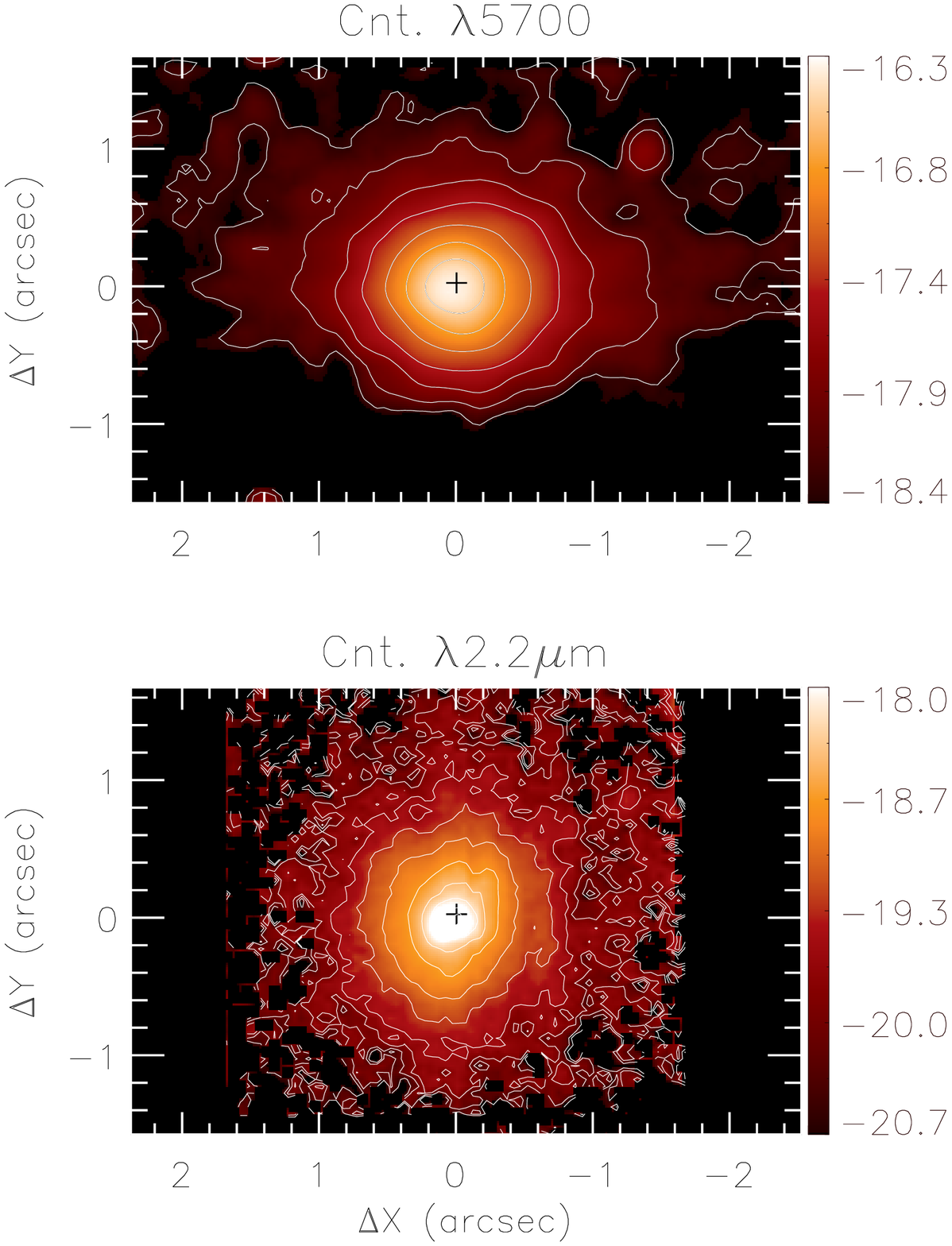}
\end{tabular}
 \caption{Left panel: large-scale image of NGC\,4395 at R-band \citep{cook}.  Right panels: optical (top) and K-band  (bottom) continuum images obtained from the GMOS and NIFS data cubes, respectively. The color bars show the fluxes in logarithmic units of erg\,s$^{-1}$\,cm$^{-2}$\,\AA$^{-1}$\,spaxel$^{-1}$. The central box shown in the left panel corresponds to the GMOS FoV.}
\label{maps}
\end{figure*}    

\begin{figure*}
\centering
\includegraphics[width=0.8\textwidth]{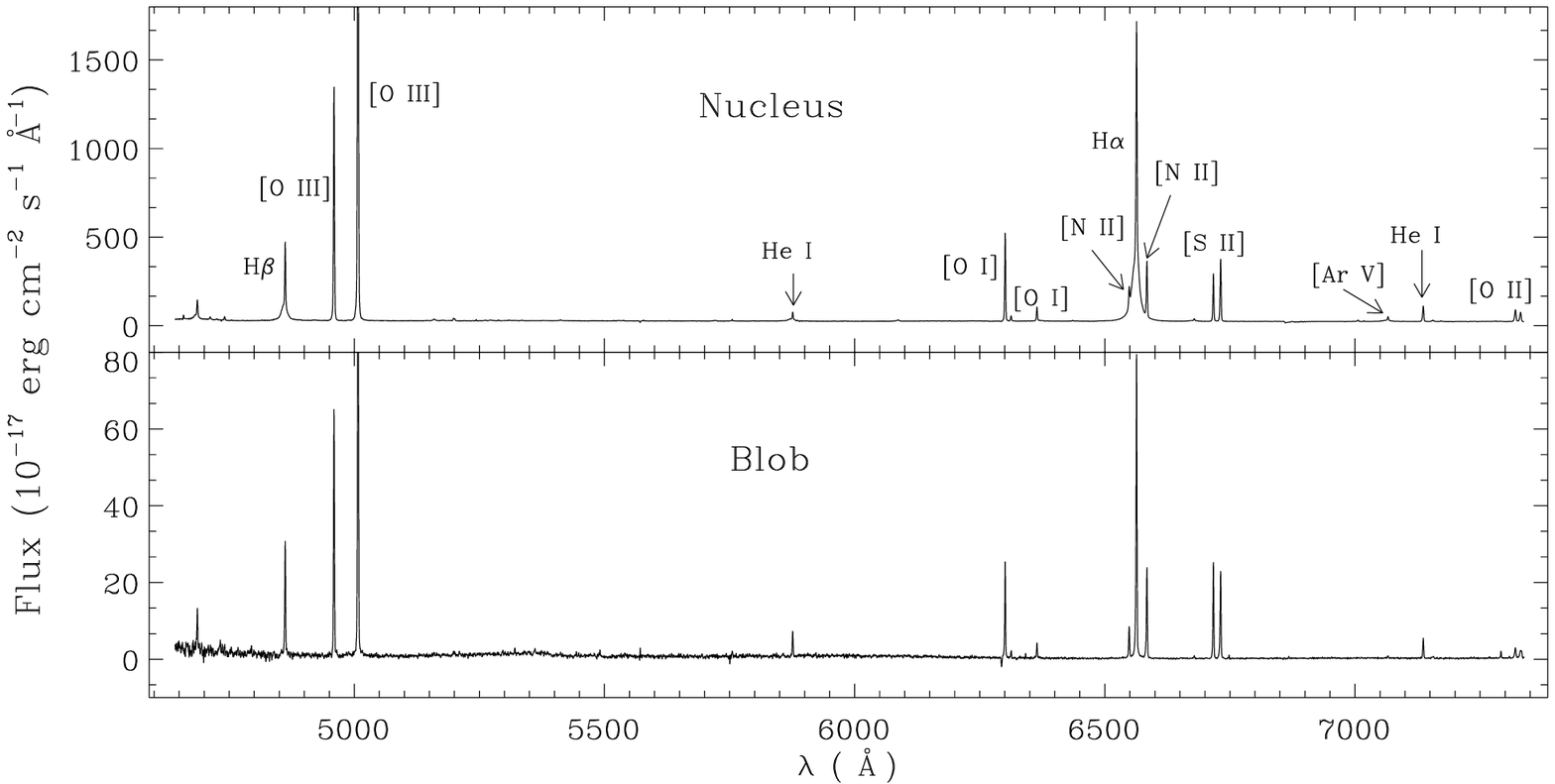}
\includegraphics[width=0.8\textwidth]{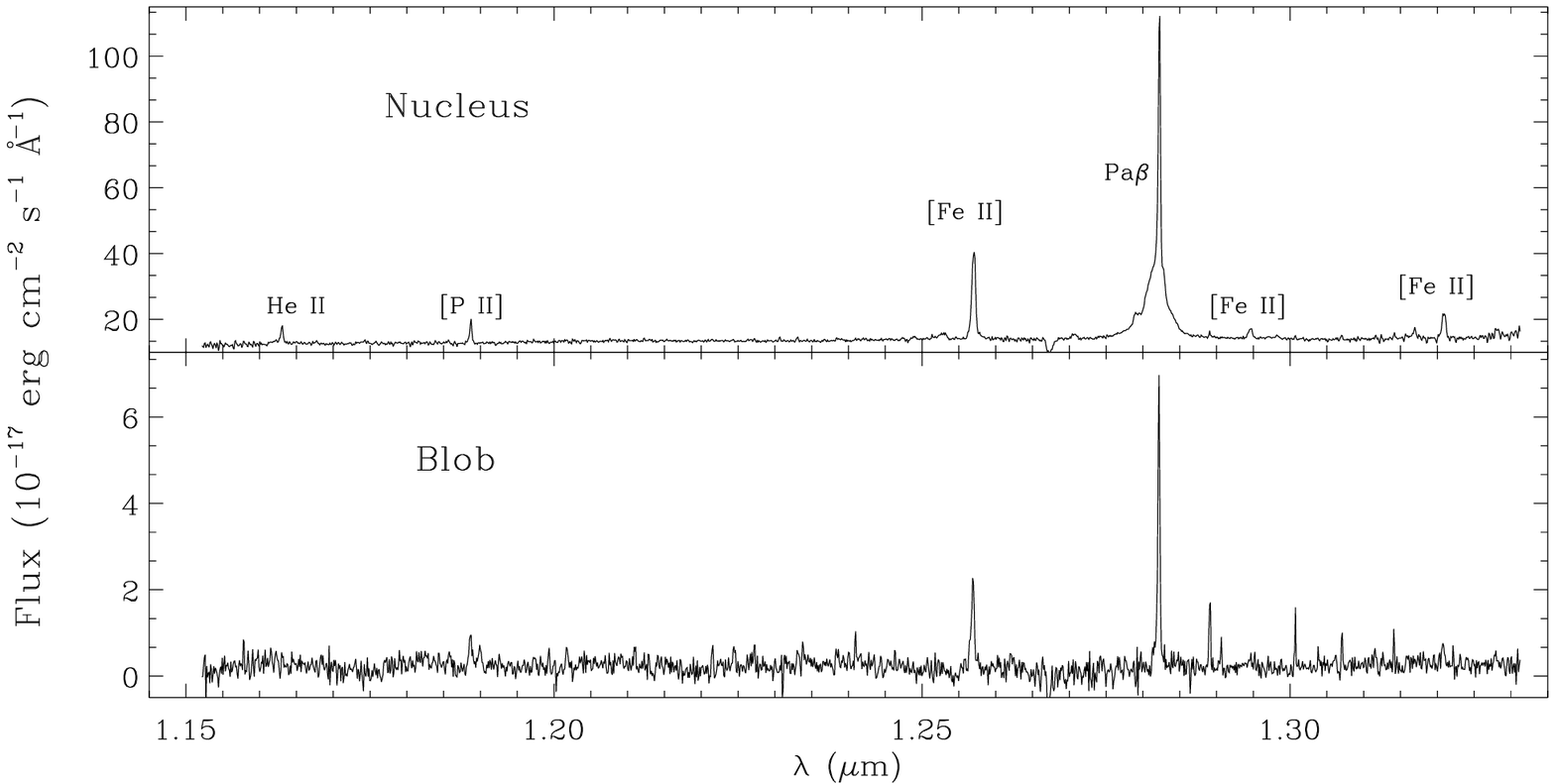}
\includegraphics[width=0.8\textwidth]{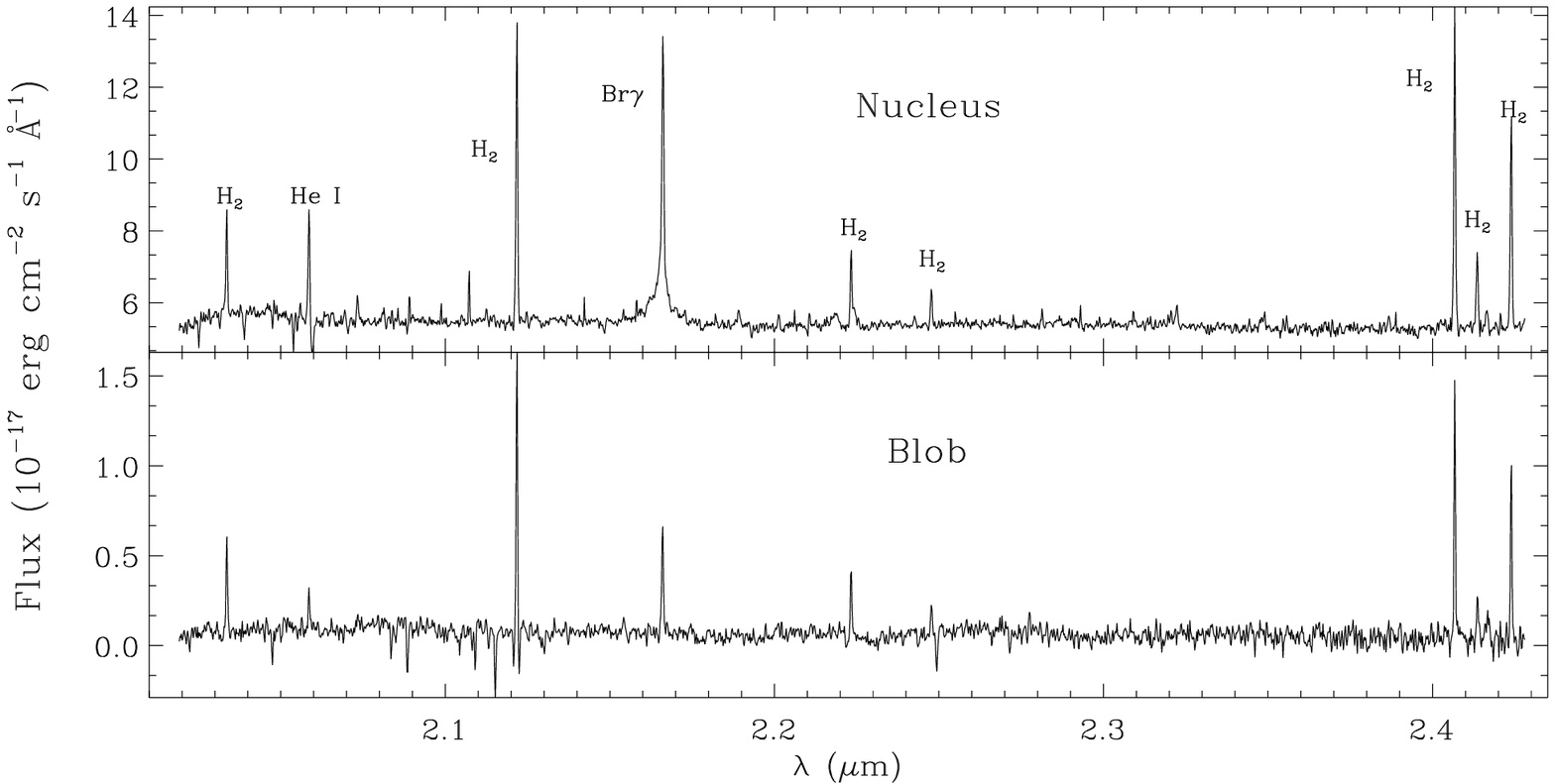}
 \caption{Optical and near-IR spectra of NGC\,4395 obtained from the GMOS and NIFS data cubes, by integrating the fluxes within a circular aperture of 0\farcs25 radius centered at the nucleus  and at  1\farcs2 west (at the blob). The strongest emission lines of each band are identified in the nuclear spectrum. The first two plots show the optical spectra, the third and forth plots show the J-band spectra and the last two show the K-band spectra. }
\label{espectros}
\end{figure*}

\section{Observations and Data Reduction}\label{obs}

Optical IFS data of NGC\,4395 were obtained with  the
 Gemini Multi-Object Spectrograph (GMOS) IFU on Gemini North Telescope, covering a spectral range 4500--7300\,\AA. The Near-IR data were collected with the Gemini Near-infrared Integral Field Spectrograph (NIFS)  at the $J$ and $K$ bands,
covering the spectral ranges 1.15--1.35\,$\mu$m and 2.01--2.43\,$\mu$m, for the J- and K-band, respectively. 
The details of the observations and data reduction procedures are discussed in the following sections.

\subsection{The optical IFU data}

NGC\,4395 was observed on 2015, June 13 with the GMOS IFU  under the program ID: GN-2015A-DD-6 (PI: Mason, R.). We used the one-slit mode, in order to obtain a spectral range that includes both the strongest emission lines in the blue (H$\beta$ and [O\,{\sc iii}]$\lambda\lambda$4959,5007) and in the red (H$\alpha$, [N\,{\sc ii}]$\lambda\lambda$6548,6584 and [S\,{\sc ii}]$\lambda\lambda$6716,31), commonly observed in active galaxies. For this configuration, the GMOS IFU science field of view (FOV)  is 5\farcs0$\times$3\farcs5. The observations were comprised of four individual exposures of 720\,s, resulting a total time of the 48 minutes with the spectra centered at $\lambda$5900\,\AA.

Data reduction was performed using the Image Reduction and Analysis Facility \citep[IRAF,][]{tody86,tody93} packages
provided by Gemini, using specific tasks developed for GMOS data in the {\sc gemini.gmos} package and followed the
standard procedure of spectroscopic data reduction \citep[see,][]{lena14}. During the reduction process the bias level was subtracted from each image, followed by flat-fielding and trimming of the spectra. The wavelength calibration was applied to the science data using as reference the spectra of arc lamps, followed by cosmic ray removal and the subtraction of the underlying sky emission. The flux calibration was done by constructing a sensitivity function using the spectra of the standard star EG\,131 observed at the same night of the galaxy's observations. 

Finally, data cubes  for each exposure were constructed with a spatial sampling of 0\farcs05$\times$0\farcs05, aligned using the peak of the continuum emission as reference and then median combined to a single cube using the {\it imcombine} {\sc iraf} task together with a sigma clipping algorithm to remove spurious features. The final data cube of NGC\,4395 covers the inner 3\farcs3$\times$4\farcs95, after the exclusion of the borders of the FoV, dominated by spurious data. The flux calibration was verified by comparing the spectrum of NGC\,4395 obtained by the Sloan Digital Sky Survey \citep[SDSS, ][]{sdss} with the GMOS spectrum, extracted within the same aperture. We found that both spectra are mutually consistent, with flux differences smaller than  5\,\% in most wavelenghts. The highest discrepancy  is seen at the red end of the GMOS spectrum, where the fluxes obtained from the GMOS data are about 20\% smaller than those obtained from the SDSS spectrum.

The spectral resolution is 1.8\,\AA\ ($\approx$90\,km\,s$^{-1}$) - derived from the full width at half-maximum (FWHM) of the Ar lamp emission lines used to perform wavelength calibration of the spectra. The angular resolution, obtained from the FWHM of the flux distribution of the broad H$\alpha$ component, is 0\farcs5, corresponding to 10 pc at the distance of NGC\,4395.

\subsection{The near-IR IFU data}

The Adaptive Optics (AO) assisted near-IR J and K-band observations of NGC\,4395 were obtained with the Near-infrared Integral Field Spectrograph  \citep[NIFS;][]{mcgregor} on the Gemini North telescope, under the programmes GN-2015A-Q-51 (PI: Mason, R.) and GN-2010A-Q-38 (PI: Seth, A., from the Gemini Observatory Archive), respectively. The total on source exposure time in the J-band was 3 hr, divided into 10 individual exposures of 480 sec each. In the K-band the total exposure time was 1.5 hr, divided into 9 exposures of 600 sec.

The data reduction for the J and K band was performed
following the same procedures used in previous works 
\citep[e.g.][]{n4051,diniz15,llp-stel}, using specific tasks contained
in the {\sc nifs} package which is part of {\sc gemini iraf} package. 
The reduction process includes the trimming of
images, flat-fielding, sky subtraction, wavelength and spatial 
distortion calibrations. The telluric absorption features were removed using the
spectrum of a telluric standard star -- of spectral type A, observed at the same nights of the galaxy's observations and the flux calibration was performed by interpolate a black body function to the spectrum
of the telluric standard star. 

For both bands, data cubes for each individual exposure were created at a spatial sampling of 0\farcs05$\times$0\farcs05, which were then median combined using a sigma clipping algorithm to eliminate the remaining cosmic rays and bad pixels, using the peak of the continuum as reference for astrometry for the distinct cubes. The final data cube for the J band covers the inner 3\farcs35$\times$ 3\farcs25 region and the final cube for the K band covers the inner 3\farcs4 $\times$4\farcs15. The FoV of both cubes is larger than the NIFS FoV due to the spatial dithering performed during the observations. 

The $J$-band observations were centered at 1.25\,$\mu$m. The spectral resolution of the data is 1.8\,\AA, measured from the FWHM of the arc lamp lines, 
corresponding to a velocity resolution of 43\,km\,$^{-1}$. 
The angular resolution is 0\farcs3, corresponding to 6\,pc at the galaxy, as 
obtained from the FWHM  of the flux distribution of the broad component of the  Pa$\beta$ emission line.
The $K$-band spectra are centered at 2.2\,$\mu$m and have a spectral resolution of  3.3\,\AA\ (45\,km\,s$^{-1}$). The angular resolution of the K-band data cube is 0\farcs2, obtained from the FWHM of the flux distribution of the Br$\gamma$ broad component emission. This value translates to a physical scale of  4\,pc at the galaxy.

In order to verify the flux calibration of the NIFS data, we had extracted nuclear spectra in the J and K band using the same aperture as those of the SDSS spectrum. As there is no spectral overlap between SDSS and NIFS data to allow a direct comparison of the spectra, we have fitted the SDSS spectra by a power law and extrapolated the resulting fit to the J and K bands. We find that the J band fluxes are consistent with the resulting fit, while the K-band spectrum shows fluxes 30\,\% smaller than the obtained from the extrapolated fit. Thus, the uncertainty due to the flux calibration is estimated to be up to 30\,\% in the K band and on average smaller than 10\% for the J and optical spectra.


\section{Results}
 
In the left panel of Figure~\ref{maps} we present an R band large-scale image of NGC\,4395 obtained with the 0.9 meter telescope of the Kitt Peak National Observatory (KPNO) \citep{cook}.  Our data cover the very inner region of this image, as indicated by the small rectangle shown in the figure to illustrate the GMOS FoV. The right panels of figure, show the optical and near-IR continuum images, obtained from the GMOS and NIFS data cubes, respectively. The optical continuum image was obtained by averaging the fluxes within a spectral window of 200\,\AA\ centered at 5700\,\AA. The near-IR (K band) image was obtained from the 
NIFS data cube using a spectral window of 100\,\AA\ centered at 2.285$\mu$m. 
At high flux levels both images show nearly circular flux contours. As the spectra of NGC\,4395 do not show any strong stellar absorption, the continuum emission is probably from the AGN. We fitted the nuclear continuum obtained by combining the optical and near-infrared spectra shown in Fig.~\ref{espectros} by a power-law function and found a spectral index of 1.1, which is consistent with an AGN origin \citep[e.g.][]{koski,kishimoto, rr09}.

\begin{figure*}
 \centering
  \includegraphics[width=0.9\textwidth]{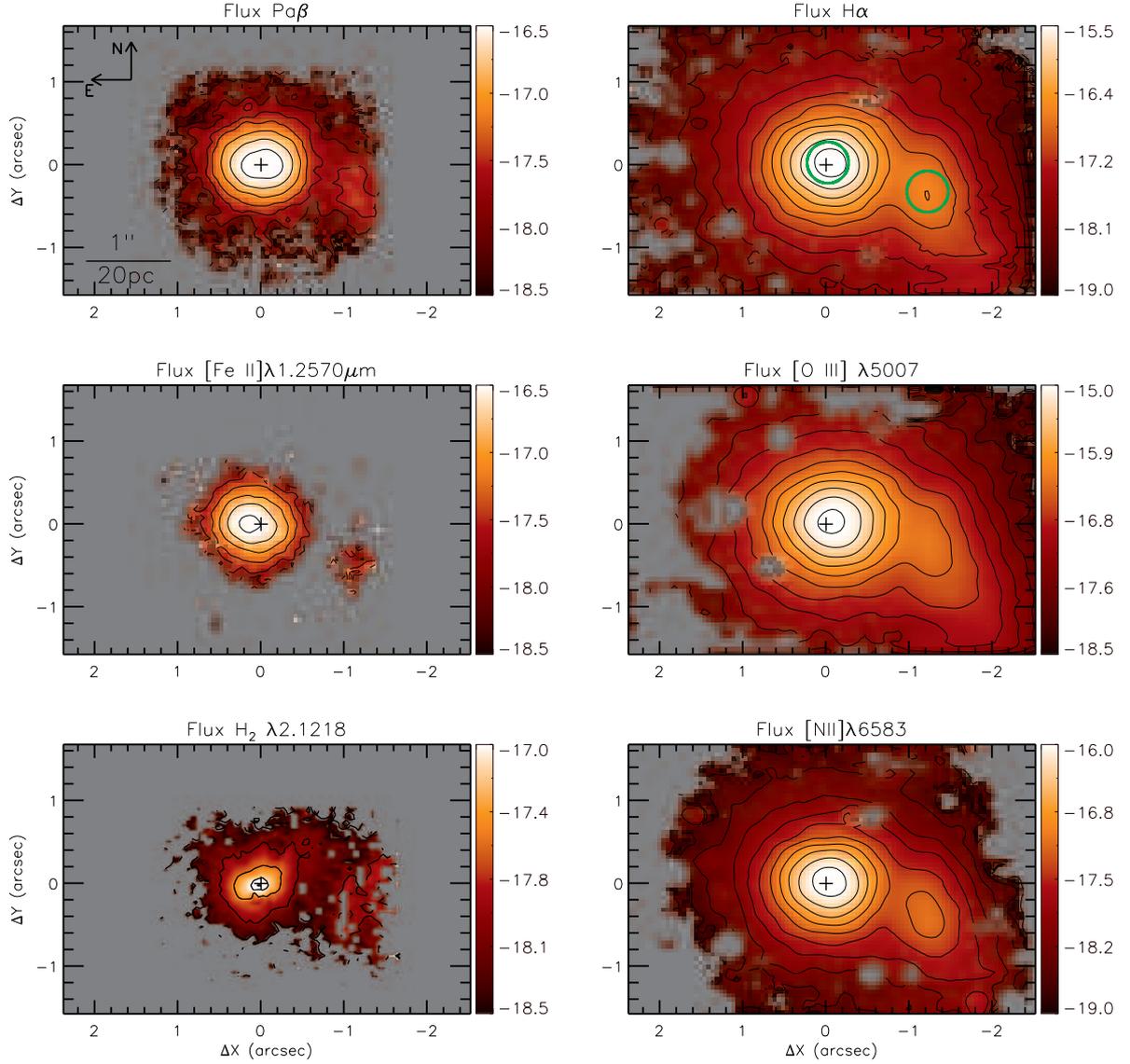} 
   \caption{Near-IR (left panels) and optical (right panels) emission-line flux distributions of NGC\,4395. The color bars show the fluxes in logarithmic units (erg\,s$^{-1}$cm$^{-2}$spaxel$^{-1}$) and the gray regions correspond to locations where the lines where not detected or non
good fits were possible. The emission line are identified in each panel and the fluxes are not corrected by extinction. The green circles shown in the H$\alpha$ map correspond to the apertures used to extract the spectra shown in Fig.~\ref{espectros}.} 
\label{flux}
\end{figure*}

In Figure~\ref{espectros} we show sample spectra of NGC\,4395 obtained from the GMOS and NIFS data. These spectra were obtained by integrating the fluxes within circular apertures of 0\farcs25 radius centered at the nucleus  and at   
 1\farcs2 west of it  -- the location of the blob seen in the HST image \citep{martini}. Here, we define the nucleus as being the location of the peak flux in the continuum. The apertures are shown as green circles in the H$\alpha$ flux map (Fig.~\ref{flux}).

The strongest emission lines are identified in the nuclear spectra. The optical nuclear spectrum of NGC\,4395 shows the \ha, H$\beta$ profiles, \oiii\,$\lambda\lambda$4959,5007,He\,{\sc i}$\lambda$5876, \oi\,$\lambda\lambda$6300,64 \nii\,$\lambda\lambda$6548,84 and [S\,{\sc ii}]\,$\lambda\lambda$6717,31 emission lines. 

Several emission lines are observed in the near-IR  nuclear spectrum of NGC\,4395. The most prominent lines are [P\,{\sc ii}]\,$\lambda\lambda$1.1471,1.1886, [Fe\,{\sc ii}]\,$\lambda\lambda\lambda$1.2570,1.2946,1.3209 and Pa$\beta$\,$\lambda$1.2821 in the J band and H$_{2}$\,$\lambda\lambda$2.1218,2.1542, Br$\gamma$\,$\lambda$2.1661 and  H$_{2}$\,$\lambda\lambda$2.2477,2.4084 in the K-band.

The observed emission lines can be used to map the flux distribution, excitation, extinction and kinematics of the molecular, low- and high-ionization gas. Below, we describe the emission-line profile fitting procedures.

The 2.3~$\mu$m CO absorption bandheads which are usu ally prominent in the spectra of nearby galaxies \citep[e.g.][]{rogerio15} were not detected in our NIFS data for NGC\,4395. Thus, we were not able to measure the stellar kinematics.
These features are expected to be detected on cold and evolved stars and their absence in the spectra of NGC\,4395 could be explained if the
continuum emission is dominated by a young-stellar population, with the spectra beeing dominated by stars hotter than F0 (see \citep{reynes}). In addtion, we do not see any important absorption feature in the spectra of NGC\,4395, thus, an possibility to explain the absence of  absorption features is that they can be diluted by the AGN continuum, as observed for nearby Seyfert galaxies \citet{rr09}.

\subsection{Emission-line profile fitting}

In order to measure the emission-line flux distributions and gas kinematics in the central region of NGC\,4395, we fitted the 
line profiles of H$\alpha$+[N\,{\sc  ii}]\,$\lambda\lambda$\,6548,6584, 
[S\,{\sc ii}]\,$\lambda\lambda$\,6717,6731, [O\,{\sc iii}]$\lambda5007$, H$\beta$, H$_{2}$\,$\lambda$\,2.1218$\mu$m,
Pa$\beta$, [Fe\,{\sc ii}]\,$\lambda$\,1.2570$\mu$m, [P\,{\sc ii}]$\lambda$1.8861$\mu$m and Br$\gamma$ at 
each spaxel over the whole FOV, by Gaussian curves using a 
modified versions of the {\sc profit} routine \citep{profit}. This routine performs a non-linear least-squares fit of the observed profiles using the MPFIT-FUN routine \citep{mark09}.  

In order to reduce the number of free parameters, we adopted the following constraints: the [N\,{\sc ii}]$+$H$\alpha$ 
emission lines were fitted by keeping tied the centroid velocities and line widths of the [N\,{\sc ii}] lines and fixing the 
[N\,{\sc ii}]$\lambda6584$/[N\,{\sc ii}]$\lambda6548$ intensity ratio to the theoretical value of 2.98, given by the ratio of their transition probabilities \citep{oster06}. As NGC\,4395 has a type 1 AGN, we included in the fit a broad
Gaussian component to represent the contribution of the Broad Line Region (BLR). As this emission is not spatially resolved by our data, the width and centroid velocity of the broad component were kept fixed for all spaxels to the values obtained from the fitting of the nuclear H$\alpha$ profile shown in Fig.~\ref{espectros}, while their amplitudes were allowed to vary to enable for smearing by the seeing. The widths and velocities of the broad components of all recombination lines were constrained to the values measured for the nuclear H$\alpha$ profile. The choice of the H$\alpha$ profile is justified as it is the strongest emission line, the broad component is well constrained for the nuclear spectrum and dust extinction seems to not play an important role for the nucleus of NGC\,4395. This is indicated by the low $E(B-V)$ values (Fig.~\ref{gmos-ratio}) and the absence of a dust emission component in the nuclear near-IR spectra (Fig.~\ref{espectros}). Similar procedures were successfully adopted in previous works \citep{brum,freitas18}.

The [S\,{\sc ii}] doublet was fitted by keeping the  centroid velocity and line width of the two lines tied, while the  [O\,{\sc iii}]\,$\lambda5007$,  [Fe\,{\sc ii}]\,$\lambda$\,1.2570$\mu$m,  
and H$_{2}$\,$\lambda$\,2.1218$\mu$m were fitted individually with all parameters free.


In all cases, the continuum emission was fitted by a linear regression, as the spectral range of each line fit was small. 
For each emission line, the fitting routine outputs a data cube with the emission-line fluxes ($F$), centroid velocity ($V$),  and velocity dispersion ($\sigma$), as well as their corresponding uncertainties and $\chi^2$ 
maps. These cubes were used to construct the two dimensional maps presented in the next sections. For the H recombination lines, all presented maps are based on the narrow-component measurements.

In Table~\ref{tab-k} we present the fluxes of optical and near-IR emission lines measured for the nucleus and the blob region using the spectra shown in Fig.~\ref{espectros}.

\begin{table}
\scriptsize
\centering
\caption{Optical and near-IR emission-line fluxes
for the nucleus and the blob region (1\farcs2 west), measured within circular aperture of 0\farcs25 using the spectra shown in Fig.~\ref{espectros}. The fluxes are not corrected by extinction and are in  units of $\rm{10^{-17}\,erg\,s^{-1}\,cm^{-2}}$. For recombination lines we labeled the broad and narrow line components as (B) and (N), respectively. }
\vspace{0.3cm}
\begin{tabular}{p{1.0cm}ccc} 
\hline
\\[-0.25cm]
$\lambda_{0}$ (\AA)  			&  ID & Nuc  & 	Blob   \\
\\[-0.25cm]
\hline
\\[-0.2cm]
4658  & 	\feiii		&$39.3\pm1.3$   & $-$  \\ \\[-0.3cm] 
4685  & 	\heii 		&$189.3\pm3.4$  & $24.1\pm2.3$  \\ \\[-0.3cm]  
4711  & 	\ariv 		&$33.1\pm2.5$   & $-$  \\ \\[-0.3cm] 
4740  & 	\ariv 		&$38.2\pm2.1$  & $5.5\pm1.5$  \\ \\[-0.3cm] 
4862  & 	\hb (N) &$809.4\pm4.0$  & $67.1\pm1.1$  \\ \\[-0.3cm] 
4862  & 	\hb (B)	& $1152.0\pm15.5$  & $-$  \\ \\[-0.3cm] 
 4959 & 	\oiii 		&$3137.0\pm2.3$ & $138.1\pm1.0$  \\ \\[-0.3cm] 
 5007 & 	\oiii 		&$9463.0\pm2.4$ & $418.6\pm0.9$  \\ \\[-0.3cm] 
 5158 & 	\fevii 		&$46.4\pm4.4$   & $2.1\pm0.9$  \\ \\[-0.3cm] 
 5199 & \feii 		&$58.5\pm3.0$  & $1.3\pm0.3$  \\ \\[-0.3cm]  
 5309 & \cav 		&$14.9\pm2.4$  & $ - $  \\ \\[-0.3cm]
 5316 & \feii 		&$4.1\pm1.1$  & $ - $  \\ \\[-0.3cm]
 5335 & \fevi 		&$13.8\pm2.1$  & $ - $  \\ \\[-0.3cm]
 5518 & \cliii 		&$6.6\pm1.8$  & $ - $  \\ \\[-0.3cm]
 5537 & \cliii 		&$7.1\pm2.2$  & $ - $  \\ \\[-0.3cm]
 5577 & \oi  		&$12.1\pm1.2$  & $ - $  \\ \\[-0.3cm]  
 5720 & \fevii 		&$27.3\pm4.1$  & $ - $  \\ \\[-0.3cm]  
 5754 & \nii  		&$19.8\pm1.4$ & $3.9\pm0.9$  \\ \\[-0.3cm]  
 5876 & 	\hei (N) &$80.9\pm1.2$  & $14.1\pm0.8$  \\ \\[-0.3cm] 
  5876 & 	\hei (B) &$172.6\pm8.0$  & $-$  \\ \\[-0.3cm] 
 6087 & \fevii	 	&$27.4\pm1.4$  &   $-$  \\ \\[-0.3cm]  
 6300 & 	\oi 		&$1111.0\pm1.0$  &  $52.6\pm1.0$  \\ \\[-0.3cm] 
 6312 &  \siii 		&$64.3\pm1.0$  &  $3.9\pm0.4$  \\ \\[-0.3cm]  
 6364 & 	\oi 		&$183.1\pm1.1$  &  $8.7\pm0.4$  \\ \\[-0.3cm] 
 6548 & 	\nii 		&$320.1\pm2.4$  &  $17.5\pm0.3$  \\ \\[-0.3cm]
 6563 & 	\ha (N)		&$3308.0\pm2.0$ &  $182.6\pm0.3$  \\ \\[-0.3cm] 
 6563 & 	\ha (B)		&$6143.0\pm18.4$ &  $-$  \\ \\[-0.3cm] 
 6584 & 	\nii 		&$868.7\pm2.1$  &  $52.8\pm0.3$  \\ \\[-0.3cm] 
 6677 &  \hei (N) 		&$22.8\pm1.1$  &  $1.6\pm0.2$  \\ \\[-0.3cm]  
 6677 &  \hei (B)  		&$74.6\pm7.4$ & $-$  \\ \\[-0.3cm]  
 6717 &  \sii 		&$659.8\pm1.0$  &  $56.9\pm0.2$  \\ \\[-0.3cm] 
 6731 &  \sii 		&$887.6\pm1.1$  &  $54.3\pm0.2$  \\ \\[-0.3cm] 
 7005 &  \arv 		&$27.6\pm1.0$  &   $-$   \\ \\[-0.3cm]  
 7064 & \hei (N)		&$46.5\pm1.0$  &  $1.7\pm0.3$  \\ \\[-0.3cm]   
 7064 & \hei (B)		&$110.4\pm6.9$  &  $-$  \\ \\[-0.3cm]   
 7135 & \ariii 		&$242.6\pm1.2$  &  $11.7\pm0.6$  \\ \\[-0.3cm]
 7155 & \feiip 		&$35.0\pm1.2$  &  $2.08\pm0.7$  \\ \\[-0.3cm]
 7172 & \feiip		&$10.9\pm1.5$ & $-$  \\ \\[-0.3cm]
 7319 & \oii 		&$211.9\pm2.6$  &  $7.7\pm0.3$  \\ \\[-0.3cm]   
 7329 & \oii		&$182.3\pm2.8$  &  $9.5\pm0.4$  \\ \\[-0.3cm]  
1.16296 & \heii (N)	&$25.2\pm2.1$  & $-$  \\ \\[-0.3cm] 
1.16296 & \heii (B) &$21.2\pm6.9$  & $-$  \\ \\[-0.3cm] 
1.18861 & \pii		&$25.2\pm1.6$  & $4.8\pm1.0$  \\ \\[-0.3cm] 
1.25702 & \feii	&$176.3\pm3.6$ & $11.3\pm1.2$  \\ \\[-0.3cm] 
1.28216 & \pab (N)				&$261.4\pm2.0$ & $22.3\pm1.0$  \\ \\[-0.3cm]
1.28216 & \pab (B)				&$754.3\pm13.2$ & $-$  \\ \\[-0.3cm]
1.29462 & \feii 	&$20.7\pm1.5$  & $0.6\pm0.3$  \\ \\[-0.3cm]  
1.32092 & \feii 	&$50.1\pm4.6$ &  $2.5\pm0.8$  \\ \\[-0.3cm]  
2.03376 & \h2 		& $16.9\pm1.6$  & $3.2\pm0.5$  \\ \\[-0.3cm]  
2.05869 & \hei (N) 		& $19.8\pm2.0$  & $1.2\pm0.4$  \\ \\[-0.3cm]  
2.05869 & \hei (B)		& $5.2\pm1.1$  & $-$  \\ \\[-0.3cm]  
2.12183 & \h2 			& $51.0\pm1.5$  & $9.2\pm0.6$ \\ \\[-0.3cm] 
2.16612 & \brg (N)				& $45.6\pm1.3$  & $3.8\pm0.4$ \\ \\[-0.3cm] 
2.16612 & \brg (B)				& $59.7\pm6.4$  & $-$ \\ \\[-0.3cm] 
2.22344 & \h2 			& $13.8\pm1.1$  & $2.3\pm0.4$ \\ \\[-0.3cm]
2.24776 & \h2 			& $6.5\pm1.0$  & $1.1\pm0.3$ \\ \\[-0.3cm] 
2.40660 & \h2 			& $53.9\pm1.4$  & $7.2\pm0.5$ \\ \\[-0.3cm]
2.41367 &  \h2 	& $15.1\pm1.6$  & $2.2\pm0.8$ \\ \\[-0.3cm]
2.42370 &  \h2 			& $38.1\pm1.4$  & $2.8\pm0.4$ \\ \\[-0.3cm]
\hline 
\end{tabular}
\label{tab-k}
\end{table}

\begin{figure}
 \centering
  \includegraphics[width=0.49\textwidth]{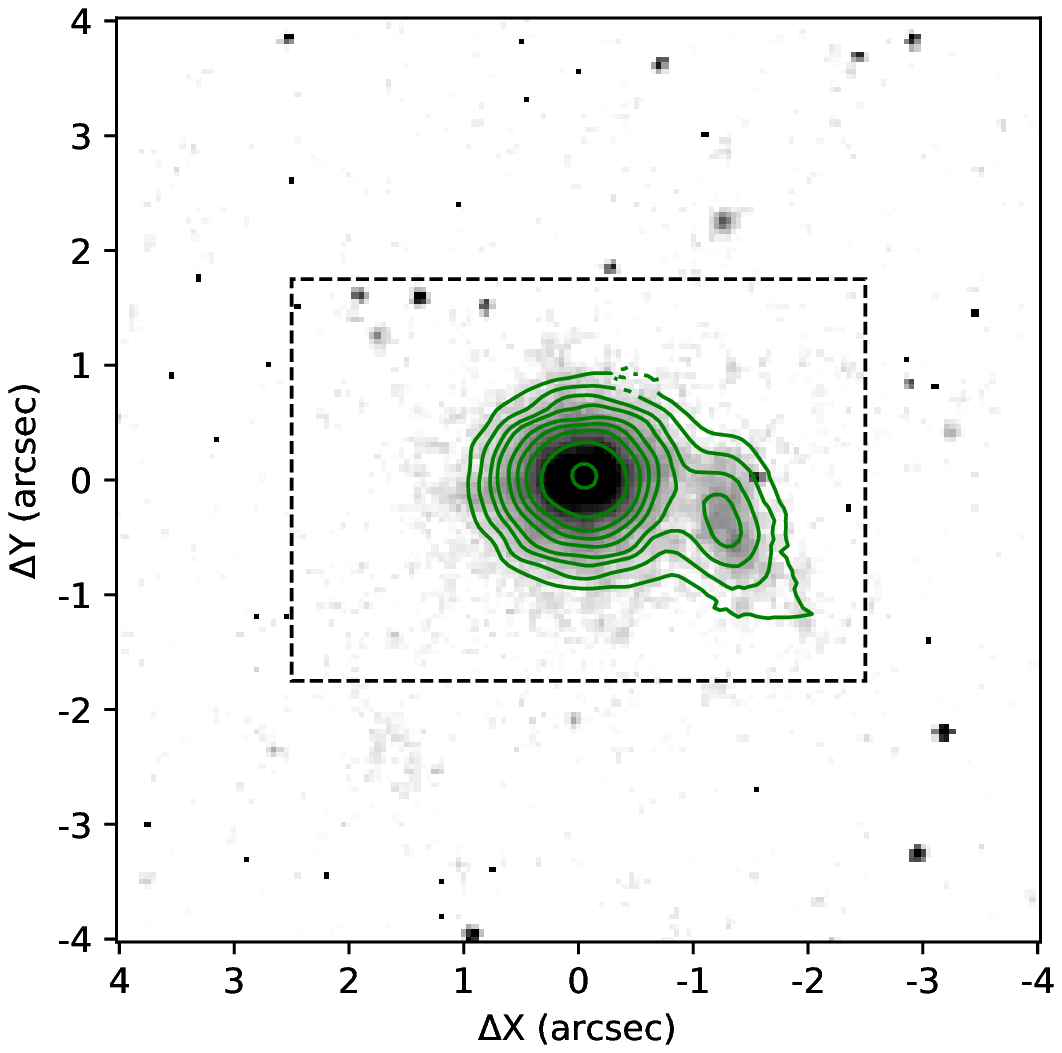} 
   \caption{Archival F606W HST image shown in grey scale. The dashed rectangle shows the GMOS IFU field of view. H$\alpha$ flux contours from the GMOS data are shown as green contours.} 
\label{hst}
\end{figure}

\subsection{Flux distributions}

The flux distributions for the emission lines Pa$\beta$, \ha, [Fe\,{\sc ii}]$\lambda$1.2570\,$\mu$m, \oiii\,$\lambda$5007\,\AA, 
H$_{2}$\,$\lambda$2.1218\,$\mu$m and \nii\,$\lambda$6584\,\AA~are presented in Fig.~\ref{flux}. Although other emission lines are observed in the spectra of NGC\,4395, we constructed maps 
for the flux distributions and kinematics only for these emission-lines, as they present the highest signal-to-noise (S/N) ratio among their species. Although the  [S\,{\sc ii}] emission-lines are detected at most locations of the GMOS FoV, we do not show the corresponding maps, as they are similar to that of \nii\,$\lambda$6584\,\AA~ line.
The color bars show the flux in logarithmic units of 10$^{-17}$erg\,s$^{-1}$cm$^{-2}$spaxel$^{-1}$. Gray regions represent masked locations where the uncertainty in the flux is larger than 30\%, and we were not able to get good fits of the line profiles due to the low S/N ratio or non-detection of the corresponding emission line. 
The left panels of Fig.~\ref{flux} present the flux maps for the near-IR lines, while the right panels present the maps for the optical lines. 
In order to make a comparison between the distinct maps, they are all shown in the same spatial scale.

The flux maps of Fig.~\ref{flux} show that all emission lines present their intensity peak at the nucleus of the galaxy, and an elongated and curved structure is seen extending to up to 2\arc\  southwest from the nucleus. This structure is co-spatial with the blob seen in the HST broad band image presented by \citet{martini}. This  correspondence is clearly seen in Fig.~\ref{hst}, where we present the HST image of NGC\,4395 obtained through the filter F606W in gray scale and show the flux contours for the H$\alpha$ flux as green contours.
As the angular resolution of the GMOS data is lower than that of NIFS data, the blob seems larger in the flux maps of the optical lines. As our continuum images (right panels of Fig.~\ref{maps}) do not show any enhanced emission to the southwest, the only plausible interpretation is that the blob seen in the HST image is due to contamination of the broad-band continuum by line emission.
 
The flux maps for \ha, \oiii~ and \nii\ are very similar. They show extended emission in almost the entire field of view (FoV). The elongated structure to the  southwest is clearly observed. 
The Pa$\beta$, [Fe\,{\sc ii}] and H$_{2}$ lines
present a more compact emission compared to the GMOS data, probably due to the higher resolution of the NIR data.


\subsection{Emission-Line Ratios}

\begin{figure*}
 \centering
  \includegraphics[width=0.9\textwidth]{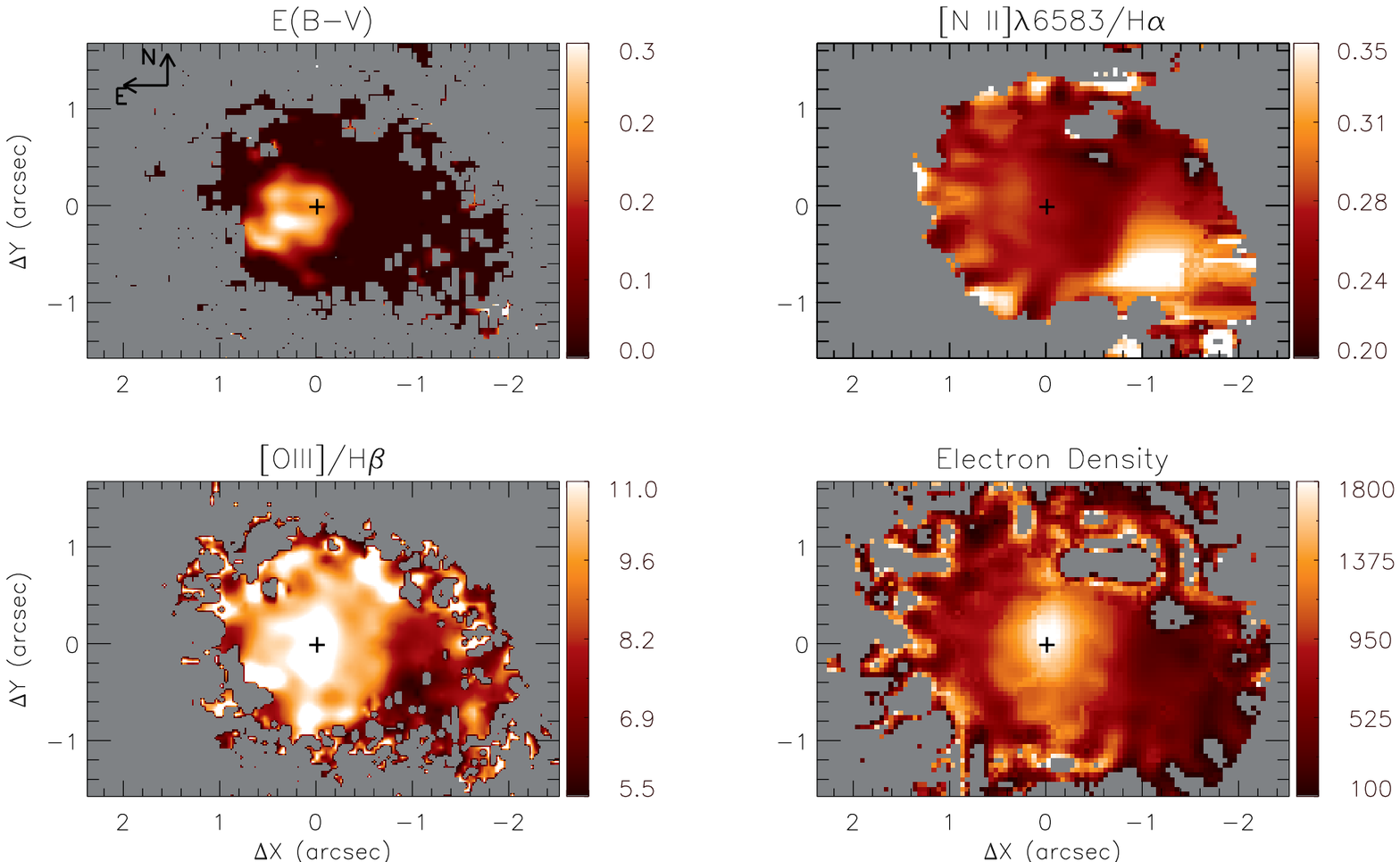} 
   \caption{Emission-line ratio maps from the GMOS data. Top-left panel: $E(B-V)$ map obtained from the H$\alpha$/H$\beta$ flux ratio. Top-right: \nii$\lambda6584$/\ha\ map. Bottom-right: \oiii$\lambda5007$/\ha\ map. Bottom-right: electron density map obtained from the [S\,{\sc ii}] lines. Density units are cm$^{−3}$. Gray regions correspond to masked locations due to the low S/N of one or both lines of the ratio.}
\label{gmos-ratio}
\end{figure*}

\begin{figure*}
  \includegraphics[width=1.02\textwidth]{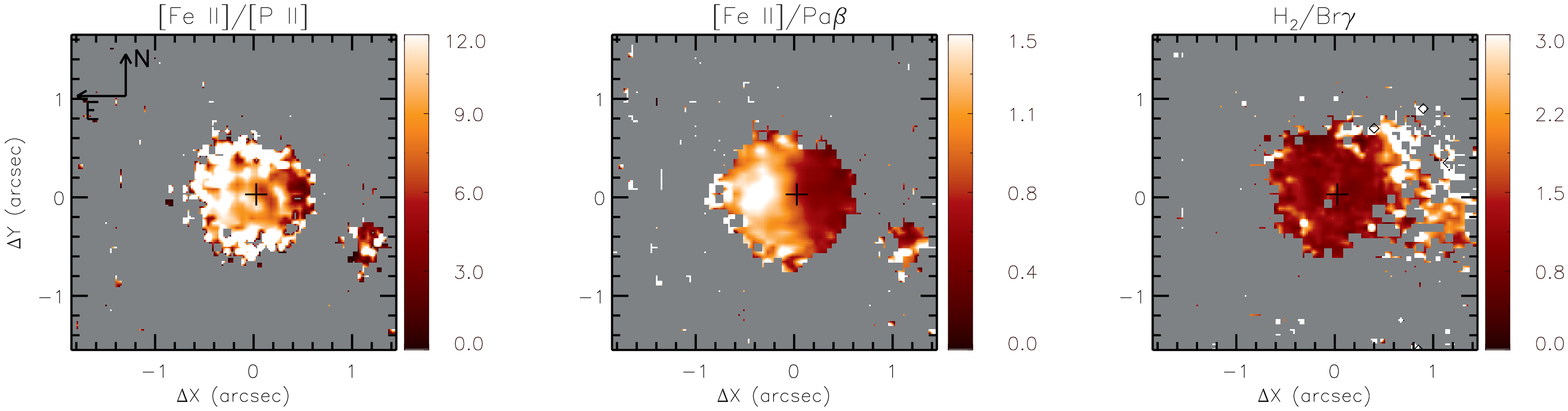}
   \caption{Emission-line ratio maps from the NIFS data. From left to right, the [Fe\,{\sc ii}]$\lambda$1.2570$\mu$m/[P\,{\sc ii}]\,$\lambda$1.8861$\mu$m,  [Fe\,{\sc ii}]\,$\lambda$1.25704$\mu$m/Pa$\beta$  and  H$_{2}$\,$\lambda$2.1218$\mu$m/Br$\gamma$ ratio maps are shown.}
\label{rat-ir}
\end{figure*}

We have used the emission-line fluxes to construct the intensity-line
ratio maps shown in Fig.~\ref{gmos-ratio} and  Fig.~\ref{rat-ir} using the optical and near-IR lines, respectively.  

\subsubsection{Optical lines}
In the top-left panel of Fig.~\ref{gmos-ratio} we show a reddening map, with values of $E(B-V)$ obtained from the \ha/H$\beta$ line ratio by    
\begin{equation}
E(B-V)\,=\,1.38\,\log {\frac{(\frac{H\alpha}{H\beta})}{2.87}}.
\label{ebveq}
\end{equation}
This equation was derived by adopting the \citet{cardelli89} reddening law and the adopting the $H\alpha/$H$\beta$ theoretical ratio for the Case B assuming a electron temperature of $Te$\,=\,10\,000\,K \citep{oster06} for the  low-density limit. At some locations the equation above returned values slightly smaller then zero, which were set as zero in the $E(B-V)$ map. 
The $E(B-V)$ map shows overall small values, with the highest ones of up to 0.3~mag being seen at the nucleus.
A small dust extinction is expected for the central region of NGC\,4395, as it does not present any signature of dust structures close to the nucleus, as revealed by the $V-H$ color map presented by \citet{martini}.

In the top-right panel of Fig.~\ref{gmos-ratio} we show the  \nii/\ha\ flux ratio map.  
It presents a narrow range of values, between 0.22 and 0.35. The highest values are seen in the blob region. Other locations show \nii/\ha\  values smaller then 0.3. The [O\,{\sc iii}]/H$\beta$ ratio map (bottom-left panel of Fig.~\ref{gmos-ratio} 
shows the highest values of up to 11 at the nucleus, while the smallest values (6--8) are observed at the blob region.

The electron density  ($N_{e}$) map is shown in the bottom-right panel of Fig.~\ref{gmos-ratio}. The  $N_{e}$ at each spaxel was derived from the intensity ratio \sii\,$\lambda$6716/$\lambda$6731 using the {\sc temden} routine of the {\sc nebular} package of the {\sc stsdas.iraf} package, assuming an electron temperature of the 10\,000\,K.  The $N_e$ values of $\approx$1800\,cm$^{-3}$ are observed at the nucleus. In the blob region, typical values are $N_e<500$\,cm$^{-3}$. 

\subsubsection{Near-IR lines}

The excitation mechanisms of the [Fe\,{\sc ii}] and H$_{2}$ emission lines can be investigated using emission-line ratio maps.
Fig.~\ref{rat-ir} shows the [Fe\,{\sc ii}]$\lambda$1.2570\,$\mu$\,m\,/\,[P\,{\sc ii}]$\lambda$1.8861\,$\mu$m (left
panel), [Fe\,{\sc ii}]$\lambda$1.2570\,$\mu$m\,/\,Pa$\beta$ (middle panel), H$_{2}$ $\lambda$2.1218$\mu$\,m/Br$\gamma$
(right panel) line flux ratio maps. These maps are shown only for the inner 3\arc$\times$3\arc. The first two line ratios are useful to investigate the origin of the [Fe\,{\sc ii}] emission and the third map can be used to investigate the H$_{2}$  emission origin \citep[e.g.][]{ardila04,ardila05,rogerio13,colina15,lamperti17}.

The [Fe\,{\sc ii}]/[P\,{\sc ii}] map shows values ranging from $\approx$3--12, with the smallest values observed west of the nucleus at distances smaller than 0\farcs8.  
In the blob region values ranging from 6--9 are observed. 

The [Fe\,{\sc ii}]/Pa$\beta$ line ratio shows values ranging from 0.3 to up to 1.0, with the highest values observed east of the nucleus and the smallest values, observed to the west. In the blob region, high values are observed.
The H$_{2}$/Br$\gamma$ line ratio shows values in the range from 0.8 to 3.0, with the highest values seen at the blob region, while at the nucleus the H$_{2}$/Br$\gamma \approx$1.2.


\subsection{Velocity and velocity dispersion maps}

\begin{figure*}
 \centering
  \includegraphics[width=0.9\textwidth]{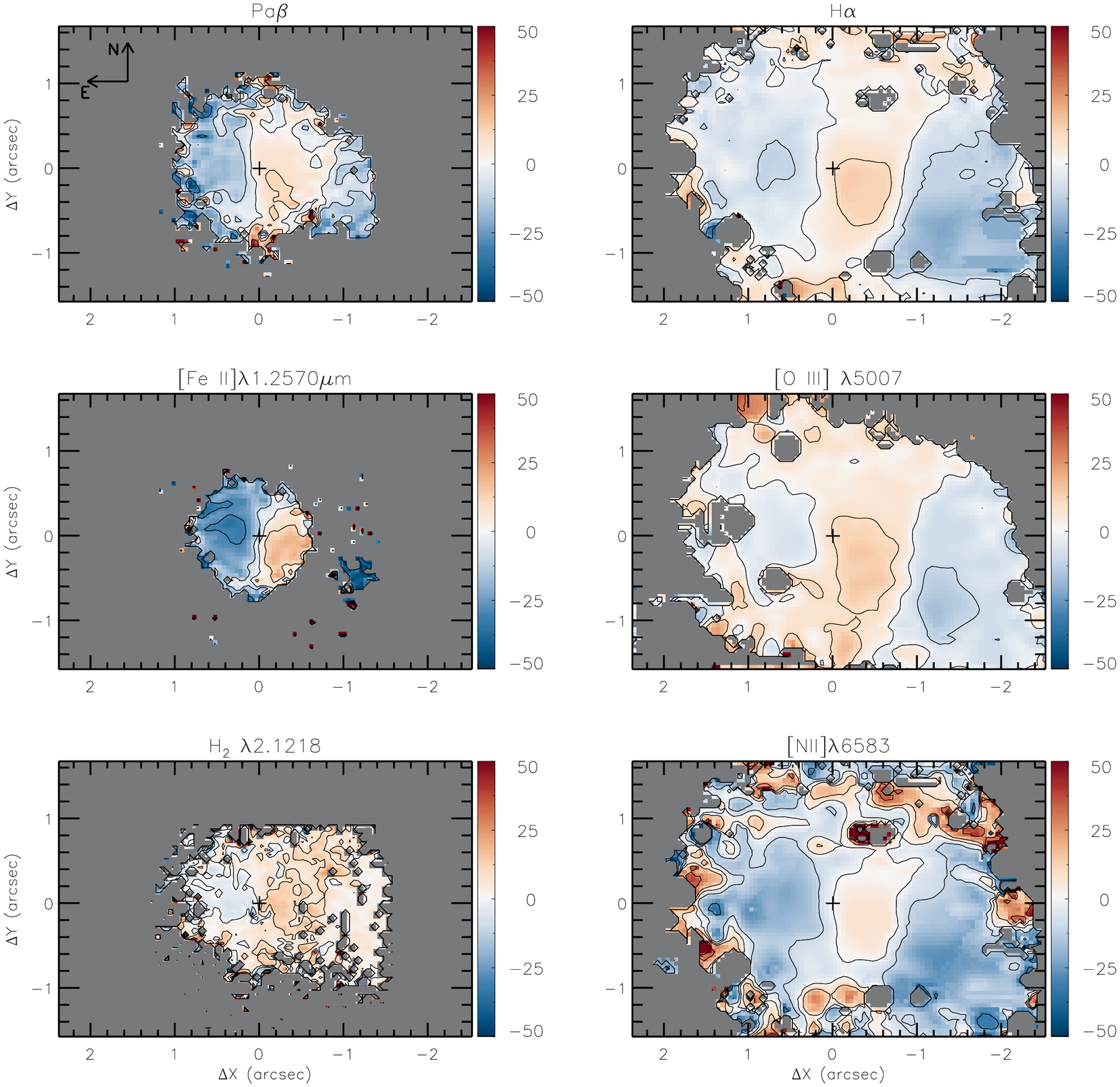} 
   \caption{Centroid velocity fields for the Pa$\beta$, \ha, [Fe\,{\sc ii}], \oiii, H$_{2}$ and \nii emission lines.  The color bar shows the velocities in units of \kms\ and gray regions represent masked locations where the S/N was not high enough to properly fit the emission lines. At most locations the uncertainties are smaller than 10\,km\,s$^{-1}$.} 
\label{vel}
\end{figure*}

\begin{figure*}
 \centering
  \includegraphics[width=0.9\textwidth]{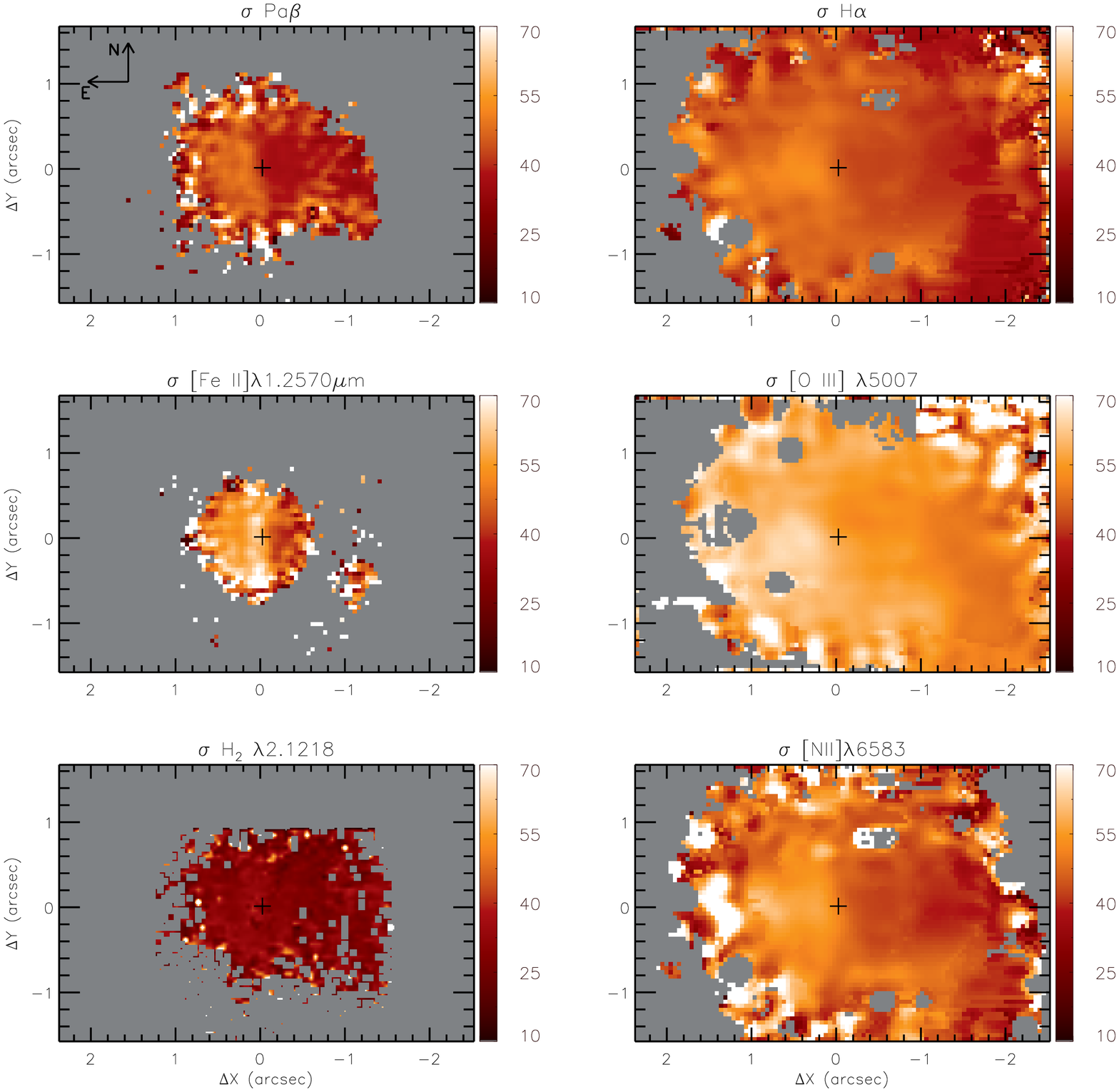} 
   \caption{$\sigma$ maps for the same emission lines of Fig.~\ref{vel}. The color bars show the $\sigma$ values in units of \kms.}
\label{sig}
\end{figure*}

In Figure~\ref{vel} we present the centroid velocity fields after subtraction of the heliocentric systemic velocity of 325\,\kms~, which was obtained 
through a model fitted to the \ha\ velocity field, as discussed in Sec.~\ref{rotmod}. 
The uncertainties in velocity are smaller than 10\,\kms at most locations. Gray 
regions in the figures represent locations where the S/N was not high enough to allow the fitting of the line profiles. 


 
The velocity fields derived for all emission lines are similar, presenting blueshifts of up to 30~\kms\ to the east side of the nucleus and similar redshifts to the west at distances smaller than 1\arc\ from the nucleus. 
In addition, blueshifts of up to 30~\kms\ are observed west of the nucleus, co-spatial with the blob region.


Figure~\ref{sig} shows the velocity dispersion maps ($\sigma$), corrected for the instrumental broadening. The uncertainties in the velocity dispersion maps are smaller than 10\,\kms at most locations. All maps show higher $\sigma$ values east of the nucleus and smaller values are observed to the west. The Pa$\beta$, H$\alpha$ and \nii\ $\sigma$ maps show the highest values of up to 60\,\kms to the east of the nucleus and lower values, down to 40\,\kms, are seen to the west. The mean $\sigma$ value at the blob region is about 55\,\kms. A similar behavior is observed in the [Fe\,{\sc ii}] and  \oiii\ $\sigma$ maps, but with overall higher $\sigma$ values. The smallest $\sigma$ values are observed in the H$_{2}$ $\sigma$ map, which shows  $\sigma<30$\,\kms\ to the west and $30<\sigma<40$\,\kms\ are observed east of the nucleus. Previous studies indicate that the velocity dispersion of the gas is similar to that of the stars \citep{greene05,ho09,kong18}. The stellar velocity dispersion  for the nucleus of NGC\,4395 ($\sigma_{\star}\approx$ 30\,\kms) was published by \citet{fili03}, measured from the Ca\,{\sc ii} absorption features. This value is similar to that observed for the H$_2$ and somewhat smaller than the $\sigma$ values for the ionized gas.


In general the H$_2$ presents the smallest $\sigma$ values, followed by the [N\,{\sc ii}] and H recombination lines and the highest $\sigma$ values are seen for the highest ionization lines of [Fe\,{\sc ii}] and [O\,{\sc iii}]. 

\section{Discussion}\label{discussion}


\subsection{Gas distributions}

The emission-line flux distributions (Fig.~\ref{flux}) reveal the presence of  an elongated to the southwest, co-spatial with a blob seen  in  HST broad band images \citep{martini,brok}. Due to the higher angular resolution, this structure is better resolved in the near-IR emission-line flux maps, while for the optical lines it appears as an elongated and curved structure. 
Considering that the our continuum images (Fig.~\ref{maps}) do not show any enhanced emission at the blob location, the most plausible interpretation is that the blob seen in the HST image is due to line emission of ionized gas within the bandwidth of the F606W HST filter. This scenario was already suggested by \citet{brok} and in Sec.~\ref{rotmod} we speculate on the origin of this blob.

The emission-line ratio maps (Figs.~\ref{gmos-ratio} and \ref{rat-ir}) show that the lowest ionization gas is located at the blob regions, while the high ionization gas is seen at the nucleus and and to the east. 
This result is consistent with the X-ray images of NGC\,4395 presented by \citep{akyuz13}, which show that the (2--12) keV emission peak at the nucleus, while lower energy (0.2--2 keV) emission is seen mainly to the east of the nucleus.

The derived electron densities for NGC\,4395 (Fig.~\ref{gmos-ratio}) show values larger than 1300~cm$^{-3}$ at the nucleus and the smallest values are observed at the blob region, which is also consistent with the observed ionization pattern. The $N_e$ value derived for the nucleus of NGC\,4395 is larger than the mean value observed for Seyfert nuclei using long-slit spectroscopy \citep[$\sim$500\,cm$^{-3}$ -- e.g.][]{dors14}. This difference may be related to the fact that the values derived from long-slit spectra are based on larger apertures, that include the circumnuclear region. Indeed, the $N_e$ map for NGC\,4395 is consistent with the range of values derived for other Seyfert galaxies using IFS \citep[e.g.][]{freitas18,kakkad18}. In a recent work, \citet{kakkad18} presented electron density maps for the inner few kiloparsec of 13 nearby active galaxies, derived from IFS and obtained as part of the {\it Siding Spring Southern Seyfert
Spectroscopic Snapshot Survey}  (S7).  For all galaxies, they found  highest values of $N_e$ at the nucleus and smaller values at the circumnuclear region. The $N_e$ map for NGC\,4395 shows a similar behavior.   


\subsection{Diagnostic Diagrams}\label{sec_diag}

To investigate the gas excitation, line-ratio diagnostic diagrams are frequently used. 
The most popular one is a plot of \oiii$\lambda5007$/H$\beta$ vs. \nii$\lambda6584$\ha\
line ratios, which is one of the Baldwin, Phillips \& Terlevich (BPT) diagrams \citep{bpt}. Several alternative diagrams for distinct spectral bands 
and new calibrations to separate distinct classes
have been proposed since the original work \citep[e.g.][]{veilleux,larkin,kewley01,kauffmann03,kewley06,reunanen02,ardila04,allen98,feltre16}. 

So far, most of the applications of diagnostic diagrams to the study of galaxies are based on single 
aperture spectra, and thus with very limited spatial information. Some recent studies have been using 
integral field spectroscopy to construct two-dimensional diagnostic diagrams 
\citep[e.g.][]{belfiore16,sanchez15,colina15,sarzi10}, which reveal the presence of extra-nuclear kiloparsec scale low ionization emission-line regions (LIERs) in both 
star-forming and quiescent galaxies. Some studies have concluded that this LIER emission may be due hot, evolved (post-asymptotic giant branch - post-AGB) stars and not due to the central ionizing source \citep{cid11,singh13,stasinska15,belfiore16}.  

\begin{figure}
 \centering
  \includegraphics[width=0.36\textwidth]{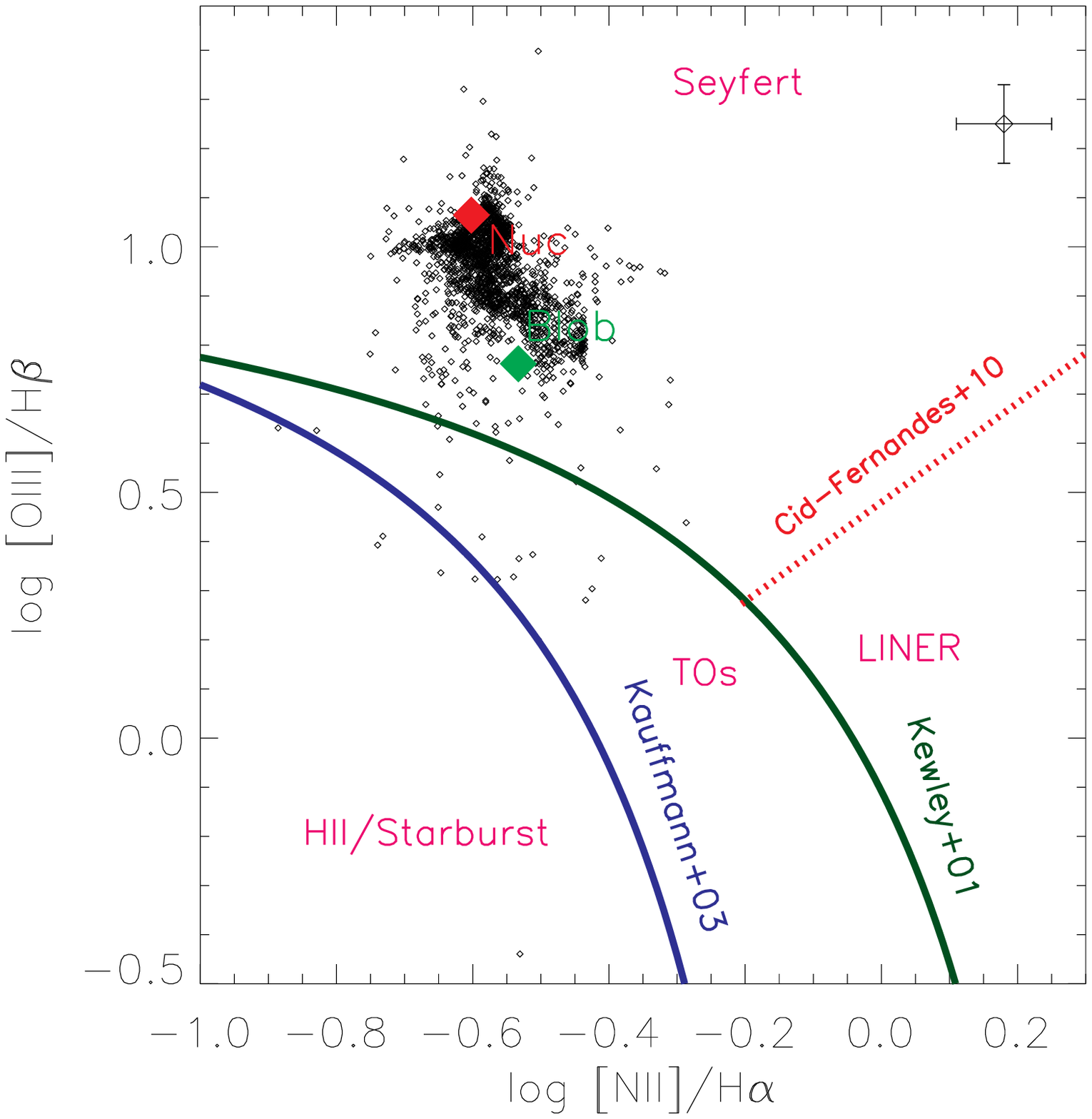} 
   \includegraphics[width=0.38\textwidth]{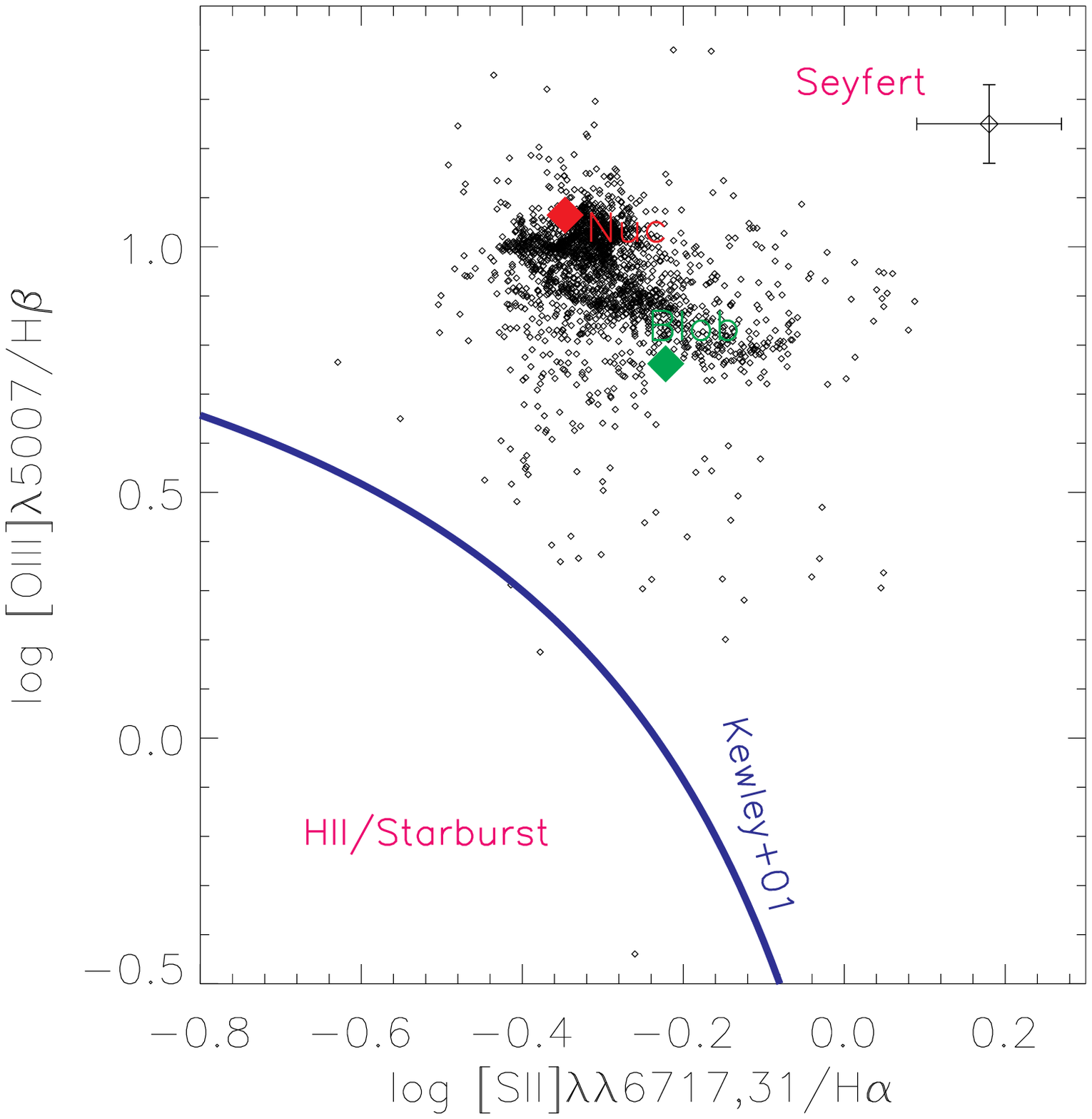}
    \includegraphics[width=0.38\textwidth]{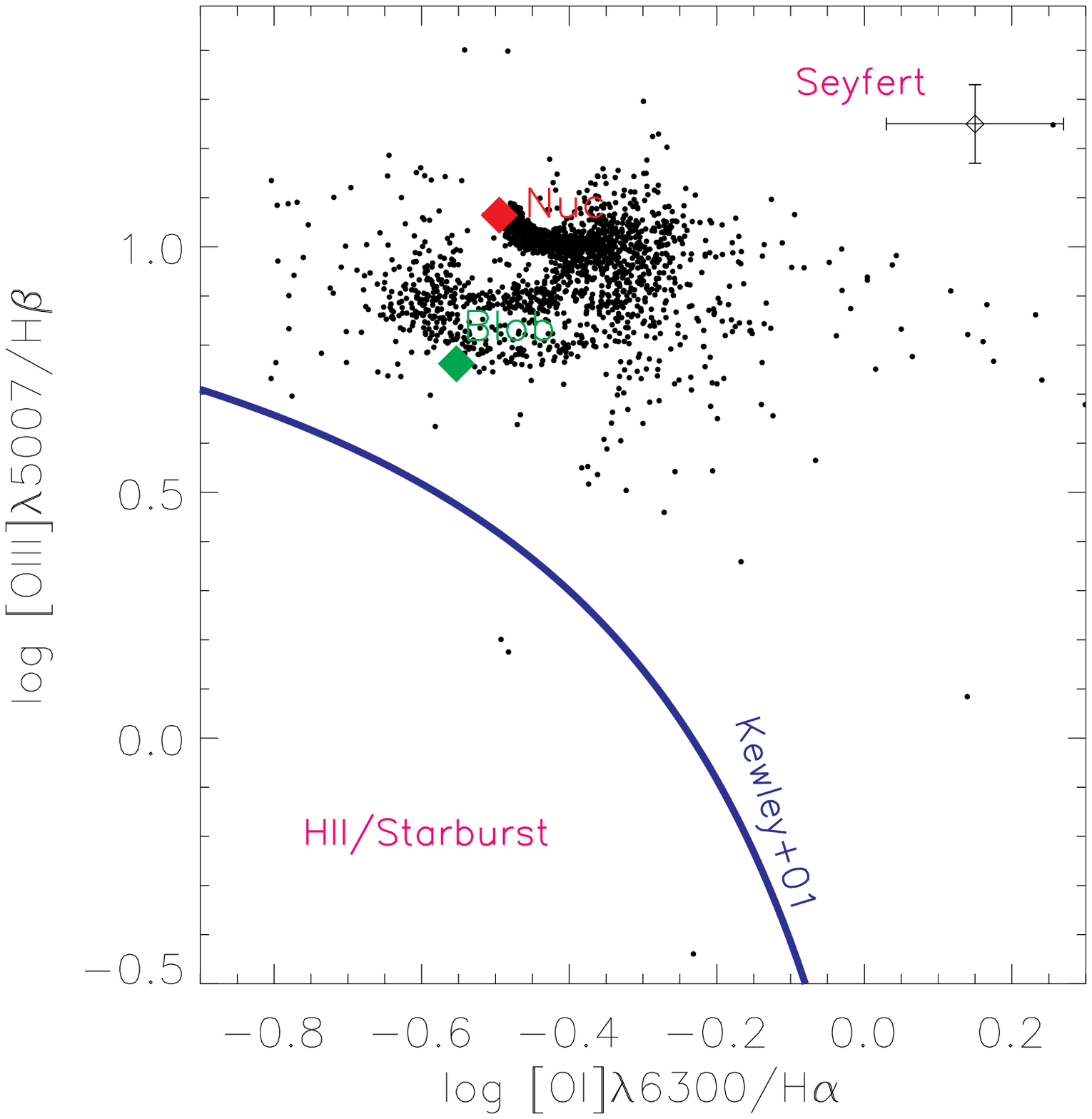}
   \caption{[O\,{\sc iii}]/H$\beta$ vs. \nii/\ha\ (top), [O\,{\sc iii}]/H$\beta$ vs. [S\,{\sc ii}]/\ha\ (middle) and [O\,{\sc iii}]/H$\beta$ vs. [O\,{\sc i}]/\ha\ (bottom) diagnostic diagrams for NGC\,4395 obtained from the GMOS data. The red and green diamonds are obtained by measuring the line ratios using the spectra shown in Fig.~\ref{espectros} for the nucleus and blob region, respectively. The uncertainties in the line flux ratios for the integrated spectra of the nucleus and blob are comparable to the size of the points and thus are not shown in the diagrams. The error bars represent the mean uncertainties in the line ratios.}
\label{bpt}
\end{figure}

In order to better map the gas excitation we have built the emission-line ratio diagnostic diagrams shown in Fig.~\ref{bpt}. Each point corresponds to one spaxel. 
The blue lines, obtained from \citet{kewley06}, are the
dividing lines between H\,{\sc ii}-type galaxies (left) and AGN (right).  The dotted line from \citet{cf10} separates the region occupied by Seyferts (above the line) and LINERs and LIERs (below the line). 
The red and green diamonds shown in this figure are for the nucleus and blob region, respectively, as obtained from the measurements of the emission-line fluxes using the spectra shown in Fig.~\ref{espectros}
As seen in Fig.~\ref{bpt}, NGC\,4395 shows intensity-line ratios over the whole GMOS FoV located in the AGN side of the [O\,{\sc iii}]/H$\beta$ vs. \nii/H$\alpha$\ diagram. However, the region occupied by NGC\,4395 is displaced to the left (smaller values of \nii/H$\alpha$), in comparison to classical Seyfert galaxies \citep[e.g.][]{cf10,freitas18}. 
This is an already known effect for AGN in dwarf galaxies, caused by the low gas metallicities observed in these objects \citep{ludwing,cedres,kraemer99}.

\begin{figure}
 \centering
  \includegraphics[scale=0.5]{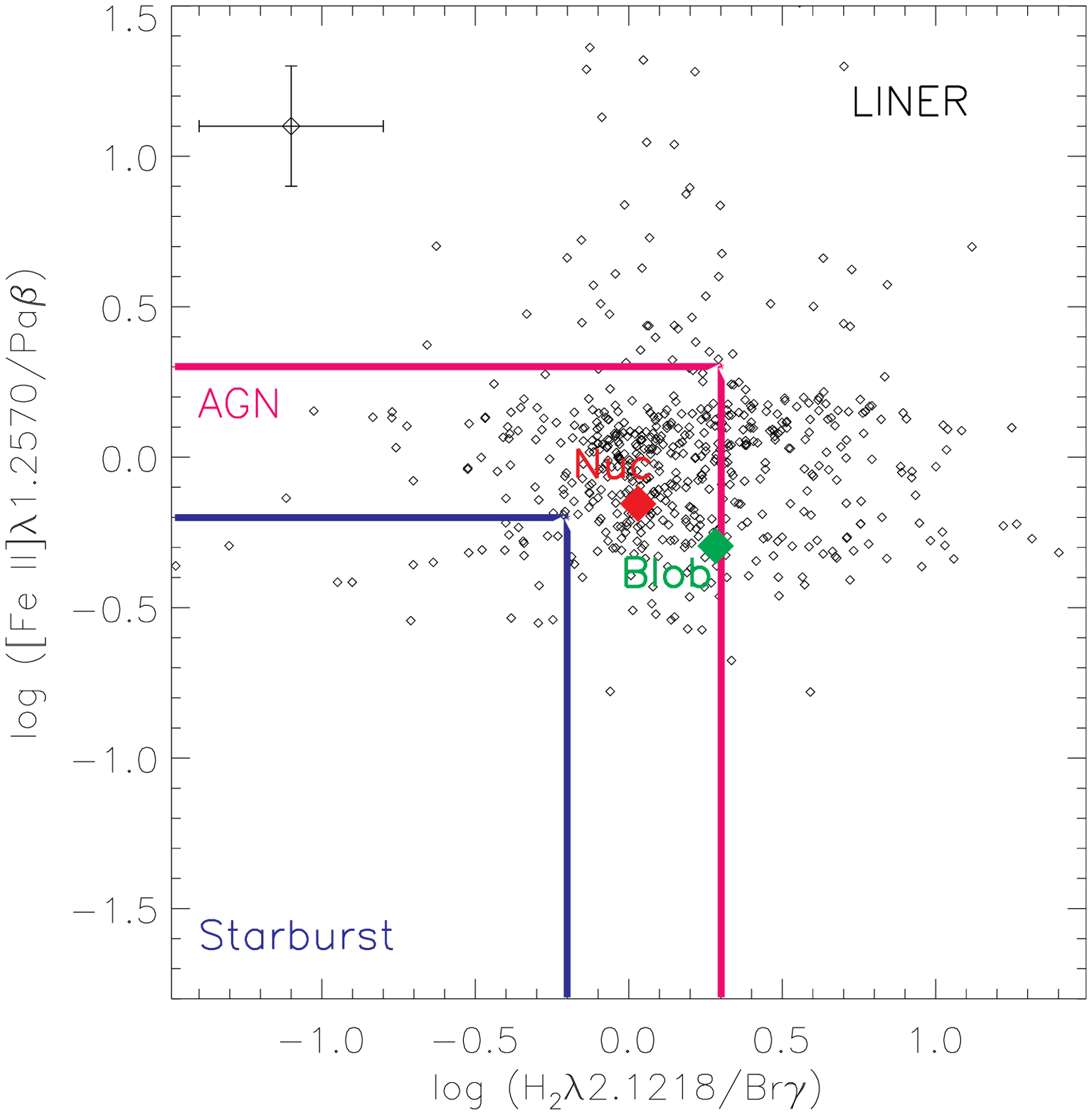} 
   \caption{[Fe\,{\sc ii}]$\lambda$\,1.2570$\mu$m\,/\,Pa$\beta$ vs. 
H$_{2}\lambda$2.1218$\mu$m\,/\,Br$\gamma$ diagnostic diagram. The red and green diagram represent line ratios for the nucleus and the blob region, as obtained from the spectra shown in Fig.~\ref{espectros}. The blue and magenta lines delimit regions with ratios typical of starbursts, Seyferts and LINERs. The uncertainties in the line flux ratios for the integrated spectra of the nucleus and blob are similar to the size of the points and thus are not shown in the plot. The error bars represent the mean uncertainties in the line ratios}
\label{bpt-ir}
\end{figure}


Further support for the results obtained using the optical line ratios can be gathered from IR diagnostic diagrams. In particular, the line flux ratios
[Fe\,{\sc ii}]$\lambda$\,1.2570$\mu$m\,/\,Pa$\beta$ vs. 
H$_{2}\lambda$2.1218$\mu$m\,/\,Br$\gamma$, shown in Fig.~\ref{bpt-ir}, also distinguish the excitation source among starburst activity, 
Seyfert activity and LINERs \citep[e.g.][]{larkin,ardila04,ardila05,mrk1066-pop}. In this diagram, Starburst/H\,{\sc ii} galaxies  occupy the  bottom left corner of the plot, with both line ratios $<$\,0.6, LINERS show both ratios larger than 2 and Seyfert nuclei are in the region between the Starburst and LINER locations, with both line ratios in the range 0.6--2  \citep[e.g.][]{ardila05,rogerio13}.

As for the optical diagram, the [Fe\,{\sc ii}]$\lambda$\,1.2570$\mu$m\,/\,Pa$\beta$ vs. 
H$_{2}\lambda$2.1218$\mu$m\,/\,Br$\gamma$  (Fig.~\ref{bpt-ir}) reveals that most locations of the central region of NGC\,4395 have flux-line ratios typical of Seyfert galaxies. At some locations, the H$_2$/Br$\gamma$ line ratio falls in the LINER region of the diagram. It should be noticed that the lines delimiting the different regions in the [Fe\,{\sc ii}]$\lambda$\,1.2570$\mu$m\,/\,Pa$\beta$ vs. 
H$_{2}\lambda$2.1218$\mu$m\,/\,Br$\gamma$  are based on fluxes measured for the nuclear spectra of galaxies. Indeed, a recent work by \citet{colina15} using IFS revealed that the H$_{2}\lambda$2.1218$\mu$m\,/\,Br$\gamma$  ratio can be higher for the extra-nuclear region of AGN-dominated sources, reaching values of up to 8.  Indeed both the nucleus and blob region present line ratios in the Seyfert region of the diagram, as obtained from the measurements of emission-line fluxes using the spectra shown in Fig.~\ref{espectros}.

We conclude that the nuclear and circumnuclear region mapped by our observations are compatible with gas ionization by the central source.

\subsection{Molecular Hydrogen and Iron Emission Origin}

The \h2~and \feii\ emission lines can be excited by thermal and non-thermal processes. The former is due to gas heating by shocks or X-rays (e.g. from an AGN), the latter is by  soft- or far-UV absorption, at dense clouds ($>10^4$\,cm$^{-3}$) \citep{sternberg89}. As discussed in Sec.~\ref{sec_diag}, all line-ratio diagnostic diagrams for NGC\,4395 suggest that the observed emission from all locations is due to gas excited by the central AGN.

We can further investigate the origin of \h2\  emission using the  H$_2\lambda2.2477/$H$_2\lambda2.1218$ line ratio \citep{mouri,reunanen02,stb09}. For values ranging from 0.1 to 0.2 the \h2~emission is due to thermal processes while values of $\approx0.55$ is attributed to non-thermal processes. Using the spectra shown in Fig.~\ref{espectros}, we obtain  H$_2\lambda2.2477/\lambda2.1218 \approx 0.13$ and $\approx 0.14$ for the nucleus and blob region, respectively. This supports an origin of the H$_2$ line emission by thermal processes, possibly associated to the AGN. The \h2~excitation temperature $T_{\rm exc}$ can be obtained using the measured fluxes of the H$_2$ emission lines together with the following expression \citep{scoville}: 

\begin{equation}
{\rm log}\left(\frac{F_{\rm i}\lambda_{\rm i}}{A_{\rm i}g_{\rm i}}\right)=
{\rm constant}-\frac{T_{\rm i}}{T_{\rm exc}},
\label{texc}
\end{equation}
where $F_{\rm i}$ is the flux of the $i\,^{th}$ \h2~line,
$\lambda_i$ is its wavelength, $A_i$ is the spontaneous emission coefficient,
 $g_i$ is the statistical weight of the upper level of the transition,
$T_i$ is the energy of the level expressed as a temperature and 
$T_{exc}$ is the excitation temperature. The Eq.~\ref{texc} assumes a ratio of transitions {\it ortho:para} of {\it 3:1} and is valid only for thermal equilibrium.

\begin{figure}
\begin{center}
    \includegraphics[width=0.48\textwidth]{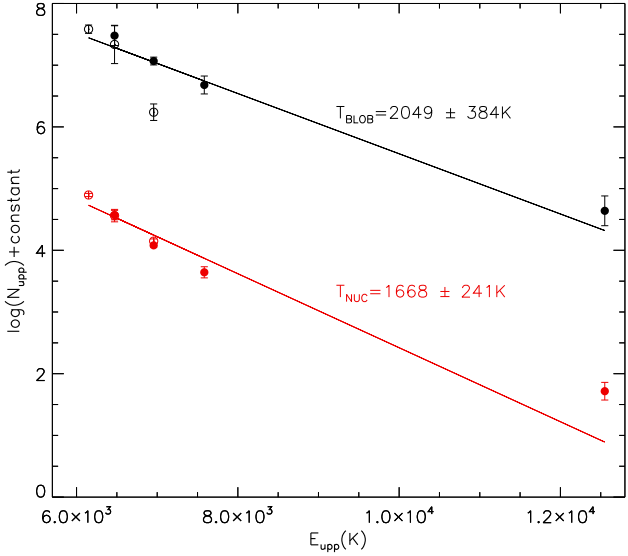}
\caption{\small Relation between 
$N_{\rm upp} = F_{\rm i}\,\lambda_{\rm i}/A_{\rm i}\,g_{\rm i}$
and $E_{\rm upp} = T_i$ for the \h2~emission lines for thermal excitation
at the nucleus and the blob. {\it Ortho (S)} transitions are shown as
filled circles and {\it para (Q)} transitions as filled circles.}
\label{tempexc}
\end{center}
\end{figure}

Figure~\ref{tempexc} shows a plot of  N$_{\rm upp}=F_i\lambda_i/A_ig_i$ (plus an arbitrary constant) vs. $E_{\rm upp} = T_i$, where filled symbols represent {\it ortho} transition and open symbols {\it para} transitions. The best linear fit is shown as a continuous line, for the nucleus (shown in red) and the blob region (shown in black). The resulting excitation temperatures are $T_{\rm NUC}=1923\,\pm\,294\,$K and $T_{\rm BLOB}=2002\,\pm\,265\,$K for the nucleus and blob, respectively. This suggests that the vibrational and rotational states of the \h2\ gas are in thermal equilibrium, supporting that \h2~excitation mechanism is due to thermal excitation by the AGN the main responsible by molecular hydrogen emission.

The \feii$\lambda1.25\mu$m/\pii$\lambda1.188\mu$m line ratio is an important tracer of the origin of the \feii\ emission \citep{oliva}. Both \feii~and \pii~have emission lines close in wavelengths and similar excitation temperatures. Moreover, their parent ions have similar ionization potentials and radiative recombination coefficients. 
 H\,{\sc ii} regions present \feii$\lambda1.25\mu$m/\pii$\lambda1.188\mu$m$\approx2$, while for supernovae remnants where the \feii\ emission is due to shocks, the above line ratio reaches values larger than 20 \citep{oliva}. Thus, values  larger than 2 for the line ratio of \feii/\pii~indicate that shocks have passed through the gas destroying the dust grains, releasing the iron in the environment. For values higher than 20, shocks are the dominant excitation mechanism \citep{oliva,stb09}.
In NGC\,4395 (see Fig.~\ref{rat-ir}), at most locations the \feii$\lambda1.25\mu$m/\pii$\lambda1.188\mu$m ratio values range from 2 to 7, with some high values (of up to 12) observed at distances larger than 0\farcs5 from the nucleus, where the uncertainty in the line ratio is high. From the flux values quoted in Tab.~\ref{tab-k}, we obtain \feii/\pii\ of $\approx6.5$ for the nucleus and $\approx3.5$ for the blob region.  These values indicate that the \feii\ emission in NGC\,4395 is mainly due gas photoionized by its AGN radiation, but some contribution by shocks cannot be discarded.

\subsection{The broad line emission}\label{broad_line}

The FWHM of the  broad component measured for the H recombination lines using the nuclear spectrum shown in Fig.~\ref{espectros} is 785$\pm$40\,km\,s$^{-1}$. This value is smaller than that derived by \citet{kraemer99}, who obtained FWHM$\sim$1500\,\kms\ for the H$\beta$ broad component using moderate  resolution (FWHM$\sim$5--8\,\AA) spectroscopy with a 3\,m telescope at the Lick Observatory. Using the spectrum of NGC\,4395 obtained by the SDSS \citep{sdss} we derive FWHM=910\,\kms for H$\alpha$ broad component, which is more similar to the value obtained from GMOS spectra. The origin of the discrepancies among the values derived for the FWHM using different observations is not clear. A possible explanation is that the discrepancies could be due to the variability of the broad line region emission, making it difficult to properly constrain the width of the broad line component when its flux is low. 
In addition, we notice that the centroid of the broad line components is blueshifted by 45\,\kms\ relative to the velocities for the narrow lines, suggesting outflows in the BLR.

 The  observed luminosity of the broad H$\alpha$ component
  can be used to estimate the AGN bolometric luminosity ($L_{\rm bol}$) as \citep{greene05b,greene07} 
\begin{equation}
L_{\rm bol}\,=\,2.34\,\times\,10^{44}\,(L_{\rm H\alpha}/10^{42})^{0.86}.
\end{equation}

Using $L_{\rm H\alpha}=(1.2\pm0.2)\times$\,10$^{38}$~erg\,s$^{-1}$, as measured from the nuclear spectrum shown in Fig.~\ref{espectros}, we obtain $L_{\rm bol}=(9.9\pm1.4)\times\,10^{40}$~erg\,s$^{-1}$. This is larger than that derived by integrating the global spectral energy distribution of the nucleus of NGC\,4395 \citep[$L_{\rm bol}=1.9\times\,10^{40}$~erg\,s$^{-1}$;][]{moran99}. 
The  AGN bolometric luminosity can also be be estimated from the  the hard X--ray luminosity ($L_X$) by \citet{ichikawa}

\begin{equation}
    {\rm log}L_{\rm bol}\,=\,0.0378({\rm log}L_{\rm X})^2-2.03{\rm log}L_{\rm X}+61.6.
\end{equation}
For NGC\,4395, the hard X--ray (14-195 keV) luminosity is $L_X\approx(6.3\pm0.6)\times10^{40}$ erg\,s$^{-1}$ as detected in the Swift/BAT survey \citep{kyuseok}. Thus, the estimated bolometric luminosity of the AGN of NGC\,4395 is $L_{\rm bol}=(4.9\pm0.4)\times\,10^{41}$~erg\,s$^{-1}$. The value for the bolometric luminosity derived from the H$\alpha$ broad component is then intermediate.

We can estimate the mass of the central SMBH by \citep{greene07}
\begin{equation}
\left(\frac{M_{\rm BH}}{\rm M_\odot}\right)=(3.0^{+0.6}_{-0.5})\times10^6\left(\frac{L_{\rm H\alpha}}{10^{42}\,{\rm erg s^{-1}}}\right)^{0.45\pm0.03}\left(\frac{\rm FWHM_{H\alpha}}{10^3 {\rm km\,s^{-1}}}\right)^{2.06\pm0.06}.
\end{equation}
Using the $H_\alpha$ luminosity and FWHM for the broad component, listed above, we obtain $M_{\rm BH}=(2.5^{+1.0}_{-0.8})\times10^5$\,M$_\odot$, which is consistent with the values previously estimated using reverberation mapping \citep[$(3.6\pm1.1)\,\times\,10^{5}$\,M$_{\odot}$; ][] {peterson} and by modeling the molecular gas dynamics \citep[$4{_{-3}^{+8}}\,\times\,10^{5}$\,M$_{\odot}$; ][]{brok}.

\subsection{Gas kinematics}\label{rotmod}


The velocity fields shown in Fig.~\ref{vel} show redshifts to the  west of the nucleus and the blueshifts to the east. This behavior suggests the presence of a disk rotation component with the line of nodes oriented approximately along the east-west direction. Another possible interpretation for the origin of this kinematic component could be the presence of a biconical outflow driven by the central AGN of NGC\,4395. However,  the velocity dispersion maps (Fig.~\ref{sig}) show small values ($\lesssim 50$\,\kms) in all locations and an enhancement of the $\sigma$ values is expected within the cone in case of the presence of outflows due to the interaction of the winds launched from the AGN  and the ambient gas \citep[e.g.][]{eso428,freitas18}. Because of that we conclude that the gas kinematics close to the nucleus is due to rotation of gas in the plane of the disk.  In addition, non-circular motions of gas are seen to the west, as indicated by the blueshifts observed at the blob region.

We fitted a simple analytical model to the \ha~ centroid velocity field of NGC\,4395, assuming that the gas has circular
orbits in a plane of the galaxy, as done in previous works \citep[e.g.][]{couto13,allan14,allan14b,lena,lena16,brum,heka17}.
The expression for the line-of-sight velocity is given by \citet{bertola}: 

\[
V_{\rm mod}(R,\Psi)=V_{s}+ \]

\begin{equation}
~~~~~\frac{AR\cos(\Psi-\Psi_{0})\sin(i)\cos^{p}(i)}{\left\{R^2[\sin^2(\Psi-\Psi_{0})+\cos^2(i)\cos^2(\Psi-\Psi_{0})]+C_o^2\cos^2(i)\right\}^{p/2}},
\label{model-bertola}
\end{equation}
where $V_{s}$ is the systemic velocity of the galaxy, {\it A} is the velocity amplitude, 
$\Psi_0$ is the position angle of the line of nodes,
{\it i} is the disc inclination in relation to the plane of the sky, $C_0$ is a concentration parameter, defined as the radius where the rotation
curve reaches 70 per cent of its velocity amplitude, the parameter {\it p} measures the slope of the rotation curve where it flattens, in the outer region of the galaxy
and it is limited between 1$\le${\it p}$\le$3/2 (for {\it p}\,=\,1
the rotation curve at large radii is asymptotically flat while
for {\it p}\,=\,3/2 the system has a finite mass), and {\it R} is the radial distance to the nucleus projected in the plane of the sky with the corresponding position angle $\Psi$.

We used an interactive data language (IDL) routine to fit the above equation to the observed \ha\ velocity field using the {\sc mpfitfun} routine \citep{mark09} to perform the non-linear least-squares fit. As all lines show similar velocity fields, we chose the \ha\  to perform the fit, as it is the strongest emission line observed at most locations of the galaxy. During the fit, the position of the kinematical center was kept fixed to the location of the peak of the continuum emission, we fixed $p=1.5$, as done in previous works \citep[e.g.][]{brum,heka17} and the  inclination of the disk was fixed to $i=37^{\circ}$, as derived by \citet{brok} constructing dynamical models for the gas velocity field measured using NIFS data. As the blueshifts observed at the blob region trace non-circular motions, we excluded this region during the fit, by excluding all spaxels west of the nucleus observed in blueshifts.

\begin{figure*}
\begin{center}
    \includegraphics[width=1.0\textwidth]{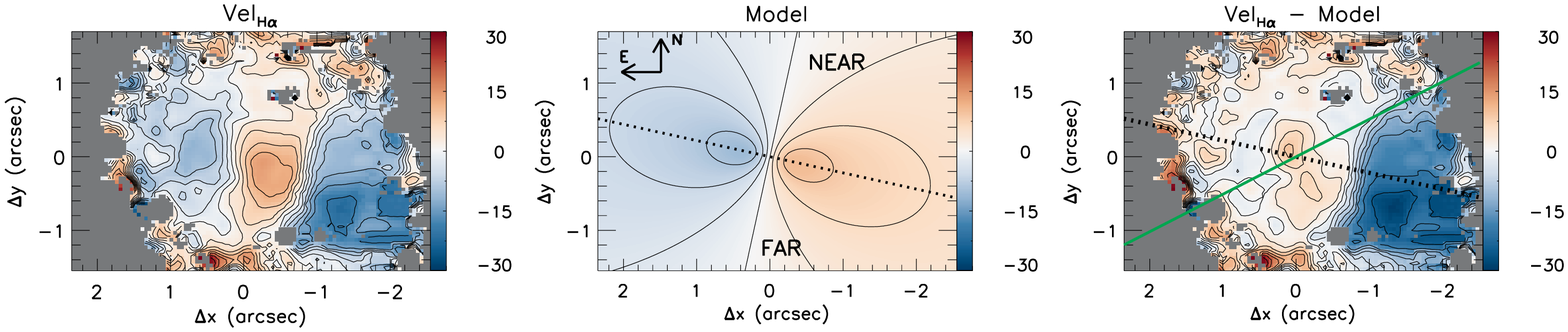} 
\caption{From left to right: \ha\ velocity field, rotating disc model for the \ha\ velocity field and residual
map (between the observed, modeled velocities). The color bars show the velocities in \kms. The dotted line displays the orientation of the line of nodes and the continuous green line shows the orientation of the photometric major axis of the galaxy from \citet{jarrett03}. The near and far side of the galaxy are indicated in the middle panel. } 
\label{model}
\end{center}
\end{figure*}

Figure~\ref{model} shows the observed \ha\ velocity field in the left panel, the best fit model in the
middle panel and the residual map, obtained by the difference  between the observed velocities and the model,  is shown
in the right panel.   
Although the velocity amplitude in the central region of NGC\,4395 is small, the rotation disk model  is a very good representation of the observed gas velocity
field,  with the exception of the blob region. 
The residuals map shows values very close to zero ($<$5\,km\,s$^{-1}$) at most other locations.

The parameters derived from the fit are: $V_s=326\pm7\,$\kms\ (corrected for the heliocentric rest frame), $A=16\pm3\,$\kms, $C_0=$0\farcs21$\pm$0\farcs03 (4.2$\pm$0.6\,pc), $\Psi_0=78^\circ\pm3^\circ$.       
 The systemic velocity resulting from the modeling is consistent with the value derived from H~{\sc i} 21~cm line emission observations of NGC\,4395 \citep[319$\pm1$\,km\,s$^{-1}$][]{haynes98}, while the orientation of the line of nodes is displaced by 39$^{\circ }$ in relation to the value obtained for the large scale disk using {\it K} band images from the 2MASS Extended Source Catalog \citep{jarrett03}. 
A possible interpretation for the difference between the kinematic position angle in the inner region and the orientation of the large scale disk is that it is due to a  kinematically distinct core (KDC), as observed for other galaxies \citep{raimundo13, raimundo17,emsellem14,davies14}. However, this interpretation should be taken with caution, as $\Psi_0$ value may not be properly constrained in our modeling due to the small field of view of the observations.

We note that the concentration parameter $C_0$ is very small, only 4.2~pc, meaning that a very compact nuclear mass distribution is driving the motions of the gas in the central region. Indeed, the observed gas velocity field shows the maximum velocity amplitude very close to the nucleus at a distance $R\sim$0\farcs5 (10~pc) from it. Assuming a simple Keplerian rotation, the mass contained within this radius can be obtained by $M=\frac{2\,R\,V_m^2}{G sin^2\,i}$, where $V_m$ is the maximum projected velocity amplitude  and $G$ is the Newton's gravitational constant. Using $V_m=25$\,\kms\ and $i=37^\circ$, we obtain $M\approx7.7\times10^5\,$M$_\odot$. 
It is well known that NGC\,4395 harbors a young (100--300\,Myr) nuclear stellar cluster  \citep{carson} and thus the derived above mass may be dominated by this stellar cluster \citep{fili93}. The mass value derived above is somewhat smaller than that found by \citet{fili03}  for the mass of the compact nucleus of NGC\,4395 (within 3.9~pc; $\sim6.2\times10^{6}\,$M$_\odot$) and is about twice the mass of the SMBH as derived by dynamical models \citep{brok} and reverberation mapping \citep{peterson}.

\subsection{The nature of the ``blob''}

In the residual map, at the blob region an excess  blueshift of $\approx$30\,km\,s$^{-1}$ is observed.  Considering that the spiral arms seen in the large scale image (Fig.~\ref{maps}) are of the trailing type, and using the observed velocity fields (Fig.~\ref{vel}) we conclude that the south-southwest is the far side of the galaxy and the north-northeast is the near side. The blob presents a small velocity dispersion, on average, of 40 \kms (Fig.~\ref{sig}) and lower ionization gas, as indicated by the line ratio maps of Fig.~\ref{gmos-ratio}, indicating that the gas is located at the plane of the galaxy. 

Two alternatives can be suggested to explain the origin of this structure. In the first case, it is a result of a minor merger of a gas rich small satellite. Similar scenarios are found for other galaxies. For example, \citet{fischer15} report an ongoing minor merger with a gas-rich dwarf galaxy inflowing towards the nucleus of Mrk\,509 and \citet{mezcua18} found that a minor merger identified in the Seyfert galaxy NGC\,5252 shows radio emission consistent with an intermediate mass black hole with mass $10^{3.5} < M_\mathrm{BH} \lesssim 2 \times 10^{6}$ M$_{\odot}$. The detection of a minor merger in a dwarf AGN opens new window to understand the feeding process of small black holes. 

Another possible scenario for the origin of the blob is that it could be a low-mass/low-metallicity cosmic gas cloud accreted onto NGC\,4395, creating a localized starburst with distinct kinematics and lowered gas-phase metallicities \citep{sanchez-almeida}. This scenario is expected in the Lambda-Cold Dark Matter context \citep{ceverino}. 

\subsection{Estimating the mass inflow rate}


We can estimate the mass of the ionized gas in the blob region as 
\begin{equation}
M_{\rm HII}=m_{p}n_{e}Vf,
\label{im}
\end{equation}
where $m_{p}$ is the proton mass, {\it n$_{e}$} is the electron density, {\it V} is the volume of the emitting region 
and {\it f} is the filling factor. The filling factor can be estimated by
\begin{equation}
L_{\rm H\alpha} \sim fN^{2}_{e}j_{\rm H\alpha}(T)V,
\label{ff}
\end{equation}
where $j_{\rm H\alpha}$=3.534$\times$10$^{-25}$\,cm$^{-3}$\,s$^{-1}$  for an electron temperature of {\it T}\,=\,10,000\,K \citep{oster06} and $L_{\rm H\alpha}$ is the H$\alpha$ luminosity emitted by a volume {\it V}. 
By replacing Eq.~\ref{ff} into Eq.~\ref{im}, we obtain

\begin{equation}
M_{\rm HII}\,=\,\frac{m_{p}L_{\rm H\alpha}}{N_{e}j_{\rm H\alpha}(T)}.
\label{mga}
\end{equation}

As the spatial resolution of our near-IR data is better than that of the GMOS data and the dust extinction in the near-IR is less important, we use the theoretical ratio between the H$\alpha$ and Br$\gamma$ ($L_{\rm Br\gamma}$) luminosities, as estimated adopting the case B of recombination for {\it T}\,=\,10,000\,K at the low-density limit \citep[$L_{\rm H\alpha}=102.1\,L_{\rm  Br\gamma}$;][]{oster06}. By replacing   $L_{\rm H\alpha}$, $m_p$ and $j_{\rm H\alpha}(T)$ in Eq.\,\ref{mga} and rewriting it, we obtain:

\begin{equation}
\left(\frac{M_{\rm HII}}{\rm M_\odot}\right)\,=2.9\times10^{19}\left(\frac{F_{\rm Br\gamma}}{\rm erg\,s^{-1}\,cm^{-2}}\right) \left(\frac{D}{\rm Mpc}\right)^2 \left(\frac{\rm cm^{-3}}{N_e}\right),
\label{mg}
\end{equation}
\noindent where $F_{\rm Br\gamma}$ is the flux of the $Br\gamma$ emission line.


The total Br$\gamma$ flux emitted by the blob region is $F_{\rm Br\gamma}\,\approx\,1.3\times10^{-16}$\,erg\,cm$^{-2}$\,s$^{-1}$, measured within a circular aperture of 0\farcs5 radius centered at the blob. As the reddening in the blob regions is very small, the correction of the above flux by extinction is negligible.  The average electron density in the blob region is 480\,cm$^{-3}$ as determined from the [S\,{\sc ii}]\,6717,31 line ratio (see Fig.~\ref{gmos-ratio} and Sect 3.3.1).
The mass of the ionized gas is, therefore $M_{\rm HII}\approx150$\,M$_{\odot}$.

As discussed above, we interpret the blueshifts seen at the blob region (24\,pc west of the nucleus) as resulting from emission of gas inflowing towards the nucleus. The observed blueshifts are  $\sim$30\,km\,s$^{-1}$ (Fig.~\ref{model}). 
Assuming that the inflows occur in the plane of the
galaxy, the deprojected inflowing velocity is $v_{\rm in}\approx49$\,km\,s$^{-1}$, using the disk inclination of $i=37^\circ$. 
If the blob is moving directly to the center with this velocity it would take $t_{\rm d}\approx\,4.8\times10^{5}$\,yr to reach the nucleus, assuming a distance of the nucleus of 1\farcs2 for the blob. The corresponding  mass-inflow rate of ionized gas is $\phi=\frac{M_{\rm HII}}{t_{\rm d}}\approx3.2\times10^{-4}\,$M$_{\odot}$yr$^{-1}$. This value is one order of magnitude smaller than those found for nearby Seyfert galaxies \citep[10$^{-3}$--  10$^{-1}$\,M$_{\odot}$yr$^{-1}$; e.g.][]{baaa}. Considering that NGC\,4395 is a dwarf low-mass galaxy, a smaller value for the inflow rate is expected.

\subsection{Molecular and ionized gas surface mass density}

The fluxes of the H$_{2}\lambda$2.12\,$\mu$m and Br$\gamma$ emission lines can be used to estimate the mass of hot molecular and ionized gas, respectively. The mass of ionized gas was estimated using Eq.~\ref{mg}.
%
%
The mass of hot molecular gas ($M_{\rm H2}$) can be estimated from \citep[e.g.][]{scoville,ngc1068-exc,llp-sample}:
\begin{equation}
\left(\frac{M_{\rm H2}}{M_\odot}\right)=5.0776\times10^{13} \left(\frac{F_{\rm H2\lambda2.1218}}{\rm erg\,s^{-1}\,cm^{-2}}\right) \left(\frac{D}{\rm Mpc}\right)^{2},
\label{molecgas}
\end{equation}
where $F_{\rm H2\lambda2.1218}$ is the flux for the H${_{2}\lambda2.1218}\mu$m emission line and a vibrational temperature of $T$\,=\,2000\,K was adopted \citep[e.g.][]{n4051,llp-sample,mrk1066-exc,stb09,astor14,astor17}.

Considering the observed $E(B-V)$ values, the correction of the fluxes of the K-band lines  due to dust extinction is small ($<$7\,\% at all locations) and thus we use the observed fluxes to derive the gas masses. 
Calculating the molecular and ionized gas masses for each spaxel and integrating over the whole NIFS FoV, we obtain 
$M_{\rm HII}\approx5.9\times10^{4}$\,M$_{\odot}$ and $M_{\rm H2}\approx2.2\,$M$_{\odot}$. To calculate $M_{\rm HII}$ we used the ${N_e}$ values derived from the [S\,{\sc ii}] lines. at each spaxel. The  $M_{\rm HII}$ and $M_{\rm H2}$ derived for NGC\,4395 are about one order of magnitude smaller than the values derived for the inner few hundred of parsecs of nearby Seyfert galaxies -- $3.5\times10^4<$M$_{\rm HII}<4.4\times10^6$\,M$_\odot$ and $50<M_{\rm H2}<2800$\,M$_\odot$ -- using the same methodology \citep{llp-sample}. However it should be noticed that the spatial coverage of the NIFS data on NGC\,4395 (tens of parsecs) is smaller than those on the Seyfert galaxies of the above study.

\begin{figure}
\centering
    \includegraphics[scale=0.55]{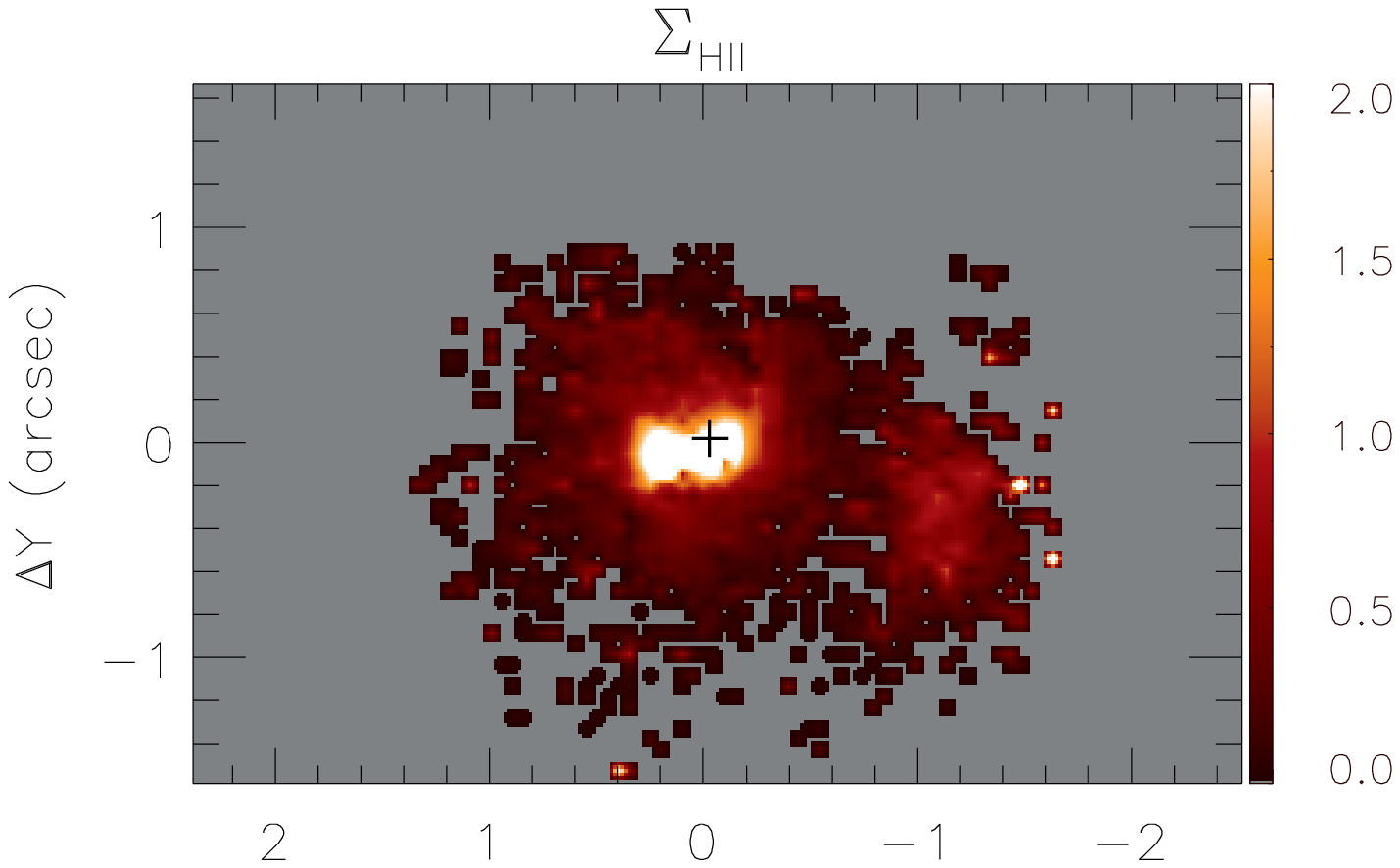} 
    \includegraphics[scale=0.55]{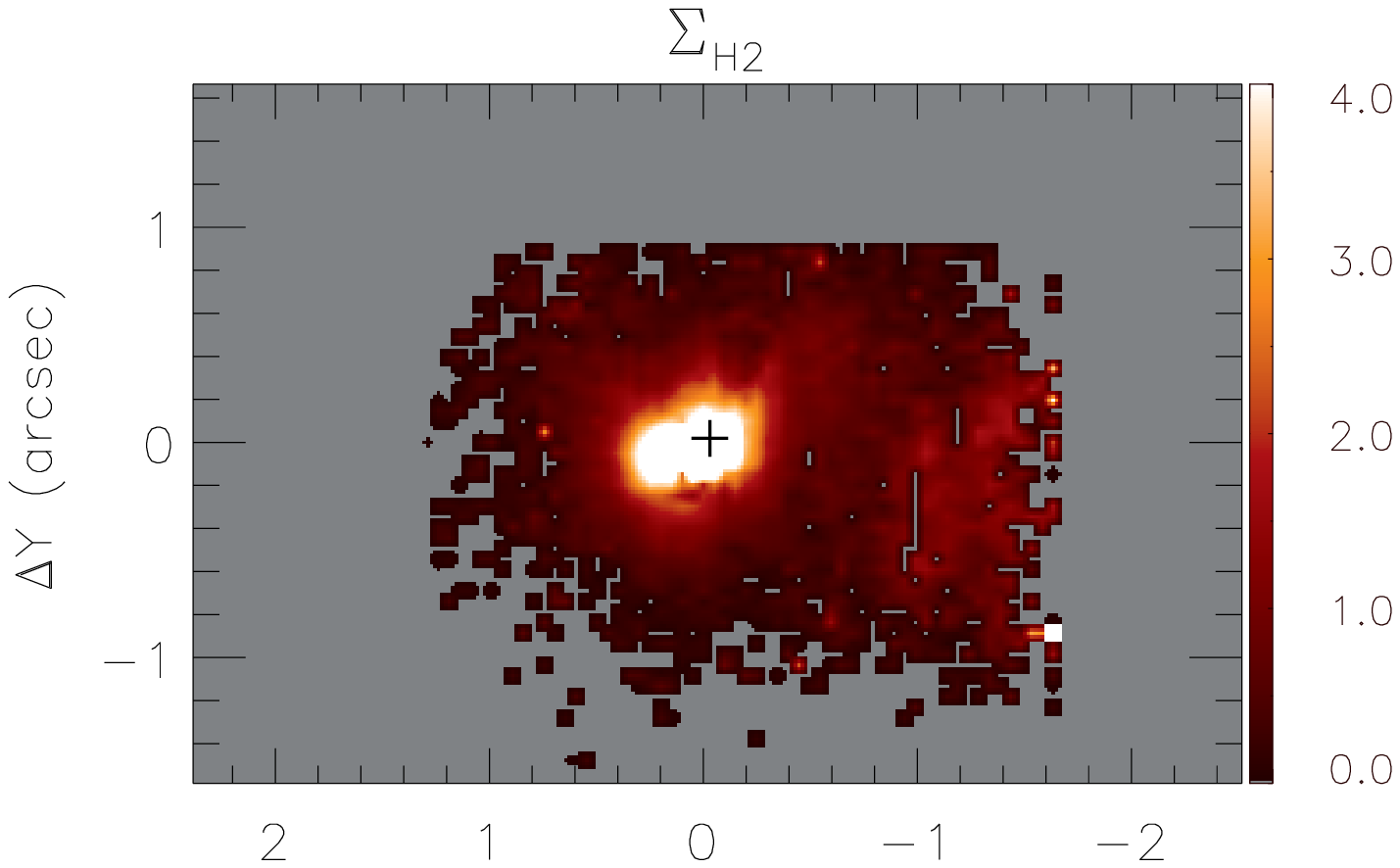} 
    \includegraphics[scale=0.55]{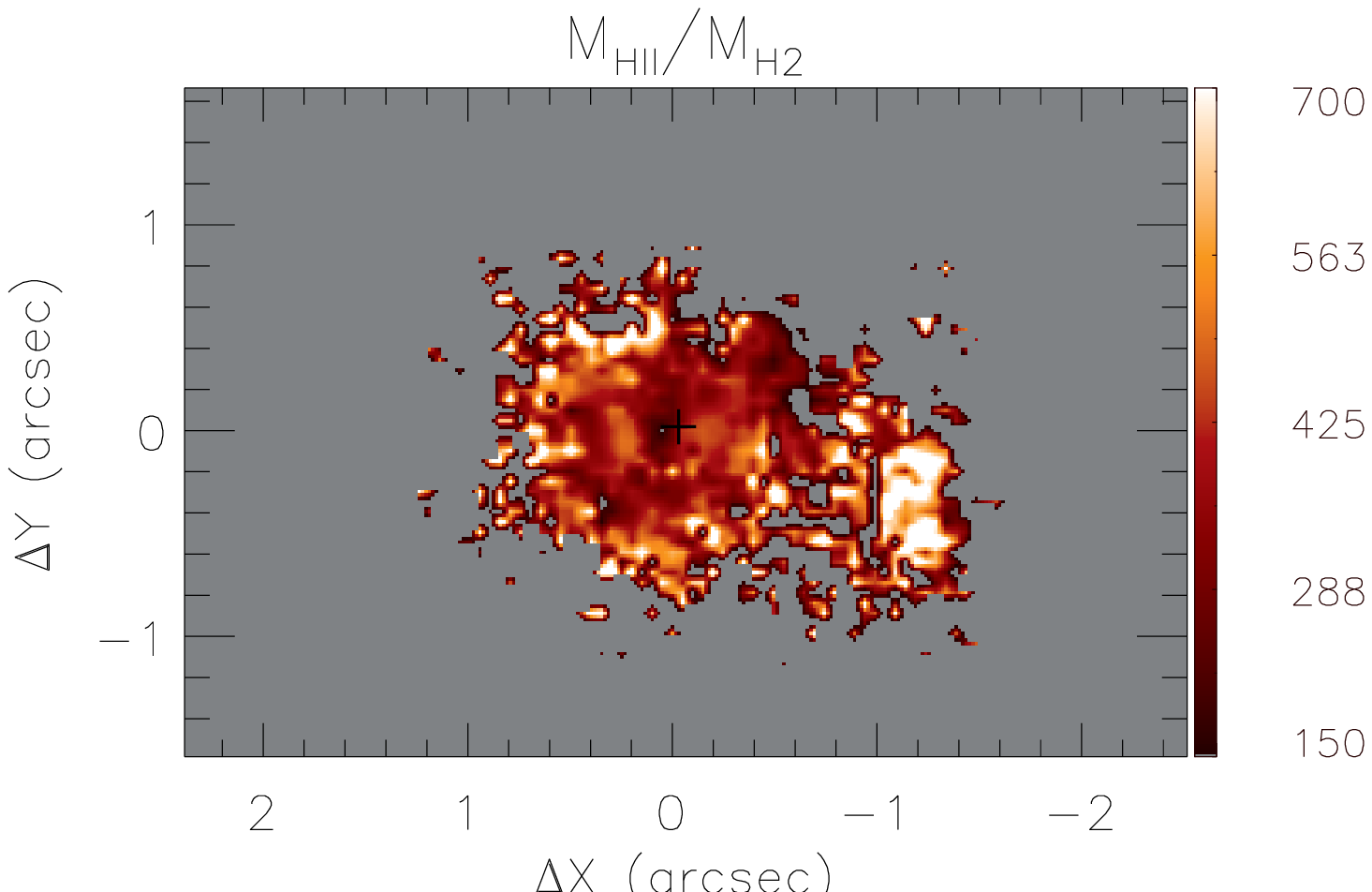} 
\caption{Surface mass density of ionized (top) and molecular (middle) gas and ratio between ionized and molecular gas masses (bottom). The color bars show the surface mass densities in units of M$_\odot$\,pc$^2$ for the ionized gas and of $10^{-3}$\,M$_\odot$\,pc$^2$ for the molecular gas.}
\label{massdensity}
\end{figure}

A better way to trace the molecular and ionized gas content is by calculating their surface mass densities, which are less sensitive to the size of the FoV. We 
calculated the ionized and molecular gas masses spaxel-by-spaxel and constructed the surface mass density ($\Sigma$) maps shown in Fig.~\ref{massdensity}. The ionized gas surface mass density ($\Sigma_{HII}$) show the highest values of $\sim$2 M$_\odot$\,pc$^{-2}$ at the nucleus, while a secondary peak is seen at the blob region with $\Sigma_{HII}$$\sim$1 M$_\odot$\,pc$^{-2}$. 
The hot molecular gas mass surface density ($\Sigma_{H2}$) shows a similar distribution, with values of  $\sim4\times10^{-3}$ M$_\odot$\,pc$^{-2}$ and $\sim1\times10^{-3}$ M$_\odot$\,pc$^{-2}$ for the nucleus and blob region, respectively. The values of $\Sigma_{HII}$ and $\Sigma_{H2}$ observed for NGC\,4395 are in the range of values typically observed for Seyfert galaxies  \citep{llp-sample}, for which $0.2<\Sigma_{\rm HII}<36$ M$_\odot$\,pc$^{-2}$ $0.2\times10^{-3}<\Sigma_{\rm H2}<14\times10^{-3}$ M$_\odot$\,pc$^{-2}$.

The bottom panel of Fig.~\ref{massdensity} shows the ratio between ionized and molecular gas masses. The smallest values of $\sim300$  are observed at the nucleus, while the highest values of up to 700 are seen at the blob region.  The smaller ratio at the nucleus  indicates that part of the molecular hydrogen is being dissociated by the radiation of the central AGN. 
\citet{llp-sample} derived ratios ranging from $\sim$200 to $\sim$8000 for nearby Seyfert galaxies and thus, the values observed for NGC\,4395 are consistent with their measurements, being similar to the lowest observed ratios. 

\subsection{Feeding the AGN of NGC~4395}

The mass accretion rate to the AGN can be estimated to compare to the accretion rate necessary to power the AGN at the nucleus of NGC\,4395, which can be derived by
\begin{equation}
 \dot{m}=\frac{L_{\rm bol}}{c^2\eta},
\end{equation},
where $L_{\rm bol}$ is the nuclear bolometric luminosity, $\eta$ is the efficiency 
of conversion of the rest mass energy of the accreted material into radiation and $c$ 
is the light speed. $L_{\rm bol}\approx\,9.9\,\times\,10^{40}$\,erg\,s$^{-1}$, as estimated in Sec.~\ref{broad_line}. Assuming $\eta\approx0.1$, which is a typical value for a ``standard'' geometrically thin, optically thick accretion disc \citep[e.g.][]{frank02}, 
we obtain a mass accretion rate of $\dot{m}\approx\,2.5\times10^{-5}~{\rm M_\odot\, yr^{-1}}$. This value is about one order of magnitude smaller than the derived mass-inflow rate, indicating that most of the gas that reach the central region of NGC\,4395 is not used to power its AGN, but most probably is consumed by star formation. This conclusion is further supported by the detection of an young nuclear star cluster in NGC\,4395 \citep{fili03,carson}. 

Considering the estimated masses of ionized and hot molecular gas ($M_{\rm HII}\approx5.9\times10^{4}$\,M$_{\odot}$ and $M_{\rm H2}\approx2.2\,$M$_{\odot}$) and the derived mass accretion rate, we conclude that the ionized gas reservoir alone  would be enough to feed the AGN of NGC\,4395 for an activity cycle of 10$^7$ -- 10$^8$ yr. In addition, it is known that the ratio between the cold and hot molecular gas masses in the central regions of galaxies is in the range $10^5-10^8$ \citep{dale2005,ms06,mazzalay13} thus the total amount of  gas in the inner 50~pc of NGC\,4395 may be at least of the order of 10$^5$ M$_\odot$. Thus, the available mass reservoir, besides feed the AGN, may also be used to form new stars. This scenario is supported by the detection of a nuclear star cluster in NGC\,4993 \citep{fili03,carson}, as well as by studies of nearby Seyfert galaxies that reveal the evidence of recent star formation \citep[e.g.][]{riffel15,llp-stel}.

\section{Conclusions}
 
  NGC\,4395 is an interesting galaxy mostly for two reasons: (i) it is very nearby so we can get a very detailed look at a Seyfert galaxy, and (ii) it is a rare, confirmed dwarf AGN that could be a local analogue of small black holes that are growing in the early universe.  We presented optical and near-IR IFS of the inner 65$\times$100\,pc$^2$ of  NGC\,4395 at angular resolutions of 0\farcs2 (4\,pc) in the {\it J}-band, 0\farcs3 (6\,pc) in the {\it K}-band and 0\farcs5 (10\,pc) in the optical.  This enabled us to estimate an extensive set of ionized and molecular gas properties, which  are summarized and interpreted as follows:

\begin{itemize}

\item The optical and near-IR emission-line flux distributions show extended emission. The line emission peaks at the nucleus but is also extended in a blob at  1\farcs2 (24\,pc)  west of the nucleus. The line ratios at all locations suggest that the gas is ionized by high-energy photons from the AGN (as opposed to shocks);   


\item The gas velocity fields are consistent with two kinematic components, one due to gas rotation at the plane of the disk of the galaxy plus another due to inflows of gas at the blob region;

\item The rotation kinematic component is well modeled by a simple analytical model, which suggests  a very compact mass distribution, with the rotation curve reaching its maximum at only 10 pc from the nucleus. We estimate a mass of $7.7\times10^5$\,M$_\odot$ inside this radius, which is about twice the estimated mass of the SMBH of NGC\,4395 and similar to the mass of the nuclear star cluster previously known; 

\item  Gas in the blob region is blueshifted by $\approx$30\,km\,s$^{-1}$ compared to the surrounding material. This is interpreted as gas flowing towards the nucleus at a rate of $\approx$ 3.2$\times10^{-4}$\,M$_{\odot}$yr$^{-1}$. The origin of the inflowing material is not clear, but it may be an ongoing minor merger of a gas-rich small galaxy, or the accretion of a low-metallicity cosmic cloud.



\item The mean surface mass density for the ionized and molecular gas are in the ranges (1--2)\,M$_\odot$pc$^{-2}$ and (1--4)$\times10^{-3}$\,M$_{\odot}$pc$^{-2}$. The ratio between the ionized and molecular gas are  $\sim$300 at the nucleus and $\sim$700 at the blob region. The gas surface densities derived for NGC\,4395 lie at the lower end of the range of values observed for typical Seyfert galaxies.

\item The mass of ionized and hot molecular gas within the inner 50 pc are $M_{\rm HII}\approx5.9\times10^{4}$\,M$_{\odot}$ and $M_{\rm H2}\approx2.2\,$M$_{\odot}$, respectively. Considering that the hot molecular gas represents only a small fraction of the total amount of molecular gas, the estimated mass reservoir would be at least 1 orders of magnitude larger than the mass needed to feed the AGN of NGC\,4395 for an AGN cycle.

\item The FWHM of broad component of the H$\,\alpha$ emission line is $\sim$785\,km\,s$^{-1}$ and the line-of-sight velocity of the BLR is blueshifted by 45\,km\,s$^{-1}$, relative to the narrow components.
 From the luminosity of the broad H$\alpha$ component, we estimate a bolometric luminosity of  $L_{\rm bol}=(9.9\pm1.4)\times\,10^{40}$~erg\,s$^{-1}$ for the AGN and estimate the mass of the SMBH as $M_{\rm BH}=(2.5^{+1.0}_{-0.8})\times10^5$\,M$_\odot$, which is consistent with previous measurements.

 NGC\,4395 differs from typical Seyfert galaxies because (a) the mass inflow rate is lower, and (b) gas inflow in Seyferts tends to happen along nuclear dust spirals and bars, whereas NGC\,4395 seems to be undergoing a minor merger or accretion event.



\end{itemize}



\section{ACKNOWLEDGMENTS}

 We thank an anonymous referee for valuable suggestions which helped to improve the paper.     This work is based on observations obtained at the Gemini Observatory, 
which is operated by the Association of Universities for Research in Astronomy, Inc., under a cooperative agreement with the 
NSF on behalf of the Gemini partnership: the National Science Foundation (United States), the Science and Technology 
Facilities Council (United Kingdom), the National Research Council (Canada), CONICYT (Chile), the Australian Research 
Council (Australia), Minist\'erio da Ci\^encia e Tecnologia (Brazil) and south-east CYT (Argentina). CB, RAR and RR acknowledge financial support from CNPq and FAPERGS. MRD thanks CAPES for financial support. The Brazilian authors thanks CAPES for financial support. LCH was supported by the National Key R\&D Program of China (2016YFA0400702) and the National Science Foundation of China (11473002, 11721303).

\label{lastpage}


\begin{thebibliography}{999}

	

\bibitem[\protect\citeauthoryear{Abolfathi et al.}{2018}]{sdss} Abolfathi, B. et al. 2018, ApJS, 235, 42
    
\bibitem[\protect\citeauthoryear{Akyuz et al.}{2013}]{akyuz13} Akyuz, A., Kayaci, S., Avdan, H., Ozel, M. E., Sonbas, E., Balman, S., 2013, ApJ, 145, 17.

\bibitem[\protect\citeauthoryear{Allen, Dopita \& Tsvetanov}{1998}]{allen98} Allen, M., G., Dopita, M., A., Tsvetanov, Z. 1998, AJ, 493, 571.

\bibitem[\protect\citeauthoryear{Baldwin,  Phillips \& Terlevich}{1981}]{bpt} Baldwin, J. A., Phillips, M. M., Terlevich, R., 1981, PASP, 93, 5.


\bibitem[\protect\citeauthoryear{Belfiore et al.}{2016}]{belfiore16} Belfiore, F., et al. 2016, MNRAS, 461, 3111.

\bibitem[\protect\citeauthoryear{Bertola et al.}{1991}]{bertola} Bertola, F., Bettoni, D., Danziger, J., 1991, ApJ, 373, 369.


\bibitem[\protect\citeauthoryear{Burtscher et al.}{2015}]{burtscher} Burtscher, L. et al. 2015, A\&A, 578, A47.




\bibitem[\protect\citeauthoryear{Brum et al.}{2017}]{brum} Brum, C., Riffel, R. A., Storchi-Bergmann, T., Robinson, A., Schnorr--M{\"u}ller, A., Lena, D., 2017, MNRAS, 469, 3405.



\bibitem[\protect\citeauthoryear{Cardelli, Clayton \& Mathis}{1989}]{cardelli89} Cardelli, J. A., Clayton, G. C. \& Mathis, J. S., 1989, ApJ, 345, 245.

\bibitem[\protect\citeauthoryear{Carson et al.}{2015}]{carson} Carson, D. J., Barth, A. J., Seth, A. C., et al. 2015, AJ, 149, 170

\bibitem[\protect\citeauthoryear{Cedr\'es \& Cepa}{2002}]{cedres} Cedr\'es, B., \& Cepa, J. 2002, \aa, 391, 809

\bibitem[\protect\citeauthoryear{Ceverino et al.}{2016}]{ceverino} Ceverino, D., S\'anchez Almeida, J., Mu\"noz Tu\"n\'on, C., Dekel, A., Elmegreen, B. G., Elmegreen, D. M., Primack, J. 2016, MNRAS, 457, 2605.
    

\bibitem[\protect\citeauthoryear{Cid Fernandes et al.}{2011}]{cid11} Cid Fernandes, R., Stasi\'nska, G., Mateus, A., Vale Asari, N., 2011, MNRAS, 413, 1687.

\bibitem[\protect\citeauthoryear{Chilingarian et al.}{2018}]{chilingarian18} Chilingarian, I. V., Katkov, I. Y., Zolotukhin, I. Y., Grishin, K. A., Beletsky, Y., Boutsia, K., Osip, D. J., 2018, arXiv180501467
    
\bibitem[\protect\citeauthoryear{Cid Fernandes et al.}{2010}]{cf10} Cid Fernandes, R., Stasi{\'n}ska, G., Schlickmann, M. S., Mateus, A., Vale Asari, N., Schoenell, W., \& Sodr{\'e} Jr, L. 2010, MNRAS, 403, 1036

\bibitem[\protect\citeauthoryear{Colina et al.}{2015}]{colina15} Colina L., Piqueras-Lopez J., Arribas S., R. Riffel, Rodriguez-Ardila, Pastoriza, M.~G., Storchi-Bergmann T., Alonso-Herrero \& Sales D., 2015, A\&A, 578, 48.

\bibitem[\protect\citeauthoryear{Cook et al}{2014}]{cook} Cook, David O., Dale, D. A., Johnson, B. D., Van Zee, L., Lee, J. C., Kennicutt, R. C., Calzetti, D., Staudaher, S. M. and Engelbracht, C., 2014, MNRAS, 445, 881

\bibitem[\protect\citeauthoryear{Couto et al.}{2013}]{couto13} Couto, Guilherme S., Storchi-Bergmann T., Axon, David J., Robinson, A., Kharb, P., and Riffel, R. A., 2013, MNRAS, 435, 2982



\bibitem[\protect\citeauthoryear{Cresci et al.}{2017}]{cresci17} Cresci, G., Vanzi, L., Telles, E., Lanzuisi, G., Brusa, M., Mingozzi, M., Sauvage, M., Johnson, K., 2017, A\&A, 604, 101.

\bibitem[\protect\citeauthoryear{Dale et al.}{2005}]{dale2005} Dale, D.~A. and Sheth, K. and Helou, G. and {Regan}, M.~W. and H{\"u}ttemeister, S., 2005, AJ, 129, 2197. 

\bibitem[\protect\citeauthoryear{Davies et al.}{2007}]{davies} Davies, R. I.; M{\"u}ller S\'anchez, F.; Genzel, R.; Tacconi, L. J.; Hicks, E. K. S.; Friedrich, S.; Sternberg, A, 2007, ApJ, 671, 1388

\bibitem[\protect\citeauthoryear{Davies et al.}{2014}]{davies14} Davies, R. I. et al. 2014, ApJ, 792, 101.

\bibitem[\protect\citeauthoryear{Den Brok et al.}{2015}]{brok} Den Brok, Mark, Seth, A. C., Barth, A. J., Carson, D. J., Neumayer, N., Cappellati, M., Debattista, V. P., Ho, L. C., Hood, C. E., McDermid, R. M. 2015, ApJ, 809, 101

\bibitem[\protect\citeauthoryear{Diniz et al.}{2015}]{diniz15} Diniz, M. R., Riffel, R. A., Stochi-Bergmann, T., Winge, C., 2015, MNRAS, 453, 1727.

\bibitem[\protect\citeauthoryear{Diniz et al.}{2017}]{diniz17} Diniz, M. R.; Riffel, R. A.; Riffel, R.; Crenshaw, D. M.; Storchi-Bergmann, T.; Fischer, T. C.; Schmitt, H. R.; Kraemer, S. B., 2017, MNRAS, 469, 3286.

\bibitem[\protect\citeauthoryear{Dong et al.}{2012}]{dong} Dong, X.-B., Ho, L. C., Yuan, W., et al. 2012, \apj, 755, 167

\bibitem[\protect\citeauthoryear{Dors et al.}{2014}]{dors14} Dors, O. L.; Cardaci, M. V., H\"agele, G. F., Krabbe, A. C., 2014, MNRAS, 443, 1291.	 

\bibitem[\protect\citeauthoryear{Emsellem et al.}{2004}]{emsellem14} Emsellem, E., Krajnovi\'c, D., Sarzi, M., 2014, MNRAS, 445, 79.	


\bibitem[\protect\citeauthoryear{Feltre, Charlot \& Gutkin}{2016}]{feltre16} Feltre, A., Charlot, S., Gutkin, J., 2016, 456, 3354. 

\bibitem[\protect\citeauthoryear{Ferrarese \& Merrit}{2000}]{ferrarese00} Ferrarese, L. \& Merrit, D. 2000, ApJ, 539, L9

\bibitem[\protect\citeauthoryear{Ferrarese \& Ford}{2005}]{ferrarese05} Ferrarese, L. \& Ford, H. 2005, SSRv, 116, 523


\bibitem[\protect\citeauthoryear{Filippenko \& Sargent}{1989}]{fili89} Filippenko, Alexei V.; Sargent, Wallace L. W., 1989, ApJL, 342, 11

\bibitem[\protect\citeauthoryear{Filippenko et al.}{1993}]{fili93} Filippenko, A. V., Ho, L. C., \& Sargent, W. L. W. 1993, ApJ, 410, L75

  \bibitem[\protect\citeauthoryear{Filippenko \& Ho}{2003}]{fili03} Filippenko, A. V., \& Ho, L. C. 2003, ApJ, 588, L13

\bibitem[\protect\citeauthoryear{Fischer et al.}{2015}]{fischer15} Fischer, T. C.; Crenshaw, D. M.; Kraemer, S. B.; Schmitt, H. R.; Storchi-Bergmann, T.; Riffel, R. A., 2015, ApJ, 799, 234.

\bibitem[\protect\citeauthoryear{Frank et al.}{2002}]{frank02} Frank J., King A., Raine D. J., 2002, Accretion Power in Astrophysics: Third Edition

\bibitem[\protect\citeauthoryear{Freitas et al.}{2018}]{freitas18} Freitas, I. C. et al., 2018, MNRAS, 476, 2760.

\bibitem[\protect\citeauthoryear{Gebhardt et al.}{2000}]{gebhardt00} Gebhardt, K. et al. 2000, ApJ, 543, L5.

\bibitem[\protect\citeauthoryear{Greene \& Ho}{2004}]{greene04} Greene, J. E., \& Ho, L. C. 2004, ApJ, 610, 722.

\bibitem[\protect\citeauthoryear{Greene \& Ho}{2005a}]{greene05} Greene, J. E., \& Ho, L. C. 2005, ApJ, 627, 721

\bibitem[\protect\citeauthoryear{Greene \& Ho}{2005b}]{greene05b} Greene, J. E., \& Ho, L. C. 2005, ApJ, 630, 122

\bibitem[\protect\citeauthoryear{Greene \& Ho}{2007}]{greene07} Greene, J. E., \& Ho, L. C. 2007, ApJ, 670, 92



\bibitem[\protect\citeauthoryear{Haynes et al.}{1998}]{haynes98} Haynes, M. P., Hogg, D. E., Maddalena, R. J., Roberts, M. S., van Zee, L., 1998, AJ, 115, 62.

         
\bibitem[\protect\citeauthoryear{Hekatelyne et al.}{2017}]{heka17} Hekatelyne, C. Riffel, R. A., Sales, D., Robinson, A., Gallimore, J., Storchi-Bergmann, T., Kharb, P., O'Dea, C., Baum, S., 2017, MNRAS, 474, 5319. 

\bibitem[\protect\citeauthoryear{Hicks et al.}{2013}]{hicks} Hicks, E. K. S.; Davies, R. I.; Maciejewski, W.; Emsellem, E.; Malkan, M. A.; Dumas, G.; M{\"u}ller-S\'anchez, F.; Rivers, A., 2013, ApJ, 768, 107

\bibitem[\protect\citeauthoryear{Ho}{2009}]{ho09} Ho, L. C., 2009, ApJ, 699, 638

\bibitem[\protect\citeauthoryear{Ichikawa et al.}{2017}]{ichikawa} Ichikawa, K., Ricci, C., Ueda, Y., Matsuoka, K., Toba, Y., Kawamuro, T., Trakhtenbrot, B., Koss, M. J., 2017, ApJ, 835, 74.

\bibitem[\protect\citeauthoryear{Jeon et al.}{2012}]{jeon} Jeon, Myoungwon; Pawlik, Andreas H.; Greif, Thomas H.; Glover, Simon C. O.; Bromm, Volker; Milosavljevi\'c , Milo\'s ; Klessen, Ralf S., 2012, ApJ, 754, 34.

\bibitem[\protect\citeauthoryear{Jarrett et al.}{2003}]{jarrett03} 
Jarrett, T. H.; Chester, T.; Cutri, R.; Schneider, S. E.; Huchra, J. P., 2003, AJ, 125, 525.

\bibitem[\protect\citeauthoryear{Kakkad et al.}{2018}]{kakkad18} Kakkad, D. et al., 2018, A\&A, 618, 6.

\bibitem[\protect\citeauthoryear{Kauffmann et al.}{2003}]{kauffmann03} Kauffmann, G. et al., 2003, MNRAS, 346, 1055.

\
\bibitem[\protect\citeauthoryear{Kewley, Heisler \& Dopita }{2001}]{kewley01} Kewley, L. J., Heisler, C. A., Dopita, M. A., 2001, ApJS, 132, 37.

\bibitem[\protect\citeauthoryear{Kewley et al. }{2006}]{kewley06}  Kewley, L. J., Groves, B., Kauffmann, G., Heckman, T., 2006, MNRAS, 372, 961.

\bibitem[\protect\citeauthoryear{Kishimoto et al.}{2007}]{kishimoto} Kishimoto, M., H\"onig, S. F., Beckert, T. \& Weigelt, G., 2007, A\&A, 476, 713.

\bibitem[\protect\citeauthoryear{Kormendy \& Ho}{2013}]{kh13} Kormendy, J., Ho, L. C., 2013, ARA\&A, 51, 511.

\bibitem[\protect\citeauthoryear{Koski}{1978}]{koski} Koski, A.T., 1978, ApJ, 223, 56.

\bibitem[\protect\citeauthoryear{Kong \& Ho}{2018}]{kong18} Kong, M.-Z., \& Ho, L. C. 2018, ApJ, 859, 116

\bibitem[\protect\citeauthoryear{Kraemer et al.}{1999}]{kraemer99}	Kraemer, S. B., Ho, L. C., Crenshaw, D. M., Shields, J. C., Filippenko, A. V., 1999, ApJ, 520, 564.

\bibitem[\protect\citeauthoryear{Kyuseok et al.}{2018}]{kyuseok} Kyuseok, O., Koss, M., Markwardt, C. B., Schawinki, K., Baumgartner, W. H., Barthelmy, S. D., Genko, S. B., Gehrels, N., Muschotzky, R., Petulante, A., Ricci, C., Lien, A., Trakhtenbrot, B., 2018 

\bibitem[\protect\citeauthoryear{Lamperti et al.}{2017}]{lamperti17} Lamperti, I., et al., 2017, MNRAS, 467, 540. 

\bibitem[\protect\citeauthoryear{Larkin et al.}{1998}]{larkin} Larkin J. E., Armus L., Knop R. A., Soifer B. T., Matthews K., 1998, ApJS, 114, 59.

\bibitem[\protect\citeauthoryear{Lena}{2014}]{lena14} Lena, D., 2014, arXiv:1409.8264

\bibitem[\protect\citeauthoryear{Lena et al.}{2015}]{lena} Lena, D., Robinson, A., Storchi-Bergmann, T., Schnorr--M{\"u}ller, A., Seelig, T., Riffel, R. A., Nagar, N. M., Couto, G.S., \& Shadler, L., 2015, ApJ, 806, 84.

\bibitem[\protect\citeauthoryear{Lena et al.}{2016}]{lena16} Lena, D. and Robinson, A.,  Storchi-Bergmann, T., Couto, G.~S.,   Schnorr-M{\"u}ller, A.,  Riffel, R.~A.,  2016, MNRAS, 459, 4485.

\bibitem[\protect\citeauthoryear{Ludwing et al.}{2012}]{ludwing} Ludwig, R. R., Greene, J. E., Barth, A. J., \& Ho, L. C. 2012, ApJ, 756, 51

	
\bibitem[\protect\citeauthoryear{Mallmann et al.}{2018}]{mallmann18} Mallmann, N. D. et al., 2018, MNRAS, 478, 5491.

\bibitem[\protect\citeauthoryear{Maraston}{2005}]{maraston05} Maraston, C., 2005, MNRAS, 362, 799.


\bibitem[\protect\citeauthoryear{Markwardt et al.}{2009}]{mark09} Markwardt C. B., 2009, in Bohlender D. A., Durand D., Dowler P., eds, ASP
Conf. Ser. Vol. 411, Astronomical Data Analysis Software and Systems XVIII. Astron. Soc. Pac., San Francisco, p. 251.

\bibitem[\protect\citeauthoryear{Martini et al.}{2003}]{martini} Martini, Paul; Regan, Michael W.; Mulchaey, John S.; Pogge, Richard W., 2003, ApJS, 146, 353.

\bibitem[\protect\citeauthoryear{Mazzalay et al.}{2013}]{mazzalay13} Mazzalay, X. et al., 2013, MNRAS, 428, 2389.

\bibitem[\protect\citeauthoryear{McGregor et al.}{2013}]{mcgregor} McGregor P. J. et al., 2003,Proc. SPIE, 4841, 1581.

\bibitem[\protect\citeauthoryear{Mezcua et al.}{2018}]{mezcua18} Mezcua, M., Kim, M., Ho, L. C., Lonsdale, C. J., 2018, arXiv:1807.03792.

\bibitem[\protect\citeauthoryear{Mouri}{1994}]{mouri} Mouri, H. 1994, ApJ, 427, 777..

\bibitem[\protect\citeauthoryear{Moran et al.}{1999}]{moran99} 	Moran, E. C., Filippenko, A..V., Ho, L. C., Shields, J. C., Belloni, T., Comastri, A., Snowden, S. L., Sramek, R. A., 1999, PASP, 111, 801.


\bibitem[\protect\citeauthoryear{M\"uller-S\'anchez et al.}{2006}]{ms06} M\"uller-S\'anchez, F., Davies R. I., Eisenhauer F., Tacconi L. J., Genzel R., Sternberg A., 2006, A\&A, 454, 481

\bibitem[\protect\citeauthoryear{M\"uller-S\'anchez et al.}{2009}]{ms09} M\"uller-S\'anchez, F., Davies, R. I., Genzel, R., Tacconi, L. J., Eisenhauer, F., Hicks, E. K. S., Friedrich, S., Sternberg, A., 2009, ApJ, 691, 739.

\bibitem[\protect\citeauthoryear{M\"uller-S\'anchez et al.}{2011}]{ms11} M\"uller-S\'anchez, F., Prieto, M. A., Hicks, E. K. S., Vives-Arias, H., Davies, R., Malkan, M.,  Tacconi, L. J., Genzel, R. , 2011, ApJ, 739, 69.

\bibitem[\protect\citeauthoryear{M\"uller-S\'anchez et al.}{2018a}]{ms18b} M\"uller-S\'anchez, F., Nevin, R., Comerford, J. M., Davies, R. I., Privon, G. C., Treister, E., 2018a, Nature, 556, 345.

\bibitem[\protect\citeauthoryear{M\"uller-S\'anchez et al.}{2018b}]{ms18a} M\"uller-S\'anchez, F., Hicks, E. K. S., Malkan, M., Davies, R., Yu, P. C., Shaver, S., Davis, B., 2018b, ApJ, 858, 48.

\bibitem[\protect\citeauthoryear{Oliva et al.}{2001}]{oliva} Oliva E. et al., 2001, A\&A, 369, L5.


\bibitem[\protect\citeauthoryear{Osterbrock \& Ferland}{2006}]{oster06} Osterbrock D. E., Ferland G. J., 2006, Astrophysics of Gaseous Nebulae
and Active Galactic Nuclei.

\bibitem[\protect\citeauthoryear{Peterson et al.}{2005}]{peterson} Peterson, Bradley M.; Bentz, Misty C.; Desroches, Louis-Benoit; Filippenko, Alexei V.; Ho, Luis C.; Kaspi, Shai; Laor, Ari; Maoz, Dan; Moran, Edward C.; Pogge, Richard W.; Quillen, Alice C., 2003, ApJ, 632, 799.

\bibitem[\protect\citeauthoryear{Raimundo et al.}{2013}]{raimundo13} Raimundo, S. I., Davies, R. I., Gandhi, P., Fabian, A. C., Canning, R. E. A., Ivanov, V. D., 2013, MNRAS, 432, 2294.

\bibitem[\protect\citeauthoryear{Raimundo et al.}{2017}]{raimundo17} Raimundo, S. I.; Davies, R. I.; Canning, R. E. A.; Celotti, A.; Fabian, A. C.; and Gandhi, P., 2017, MNRAS, 464, 4227. 

\bibitem[\protect\citeauthoryear{Reines et al.}{2011}]{reines11} Reines, Amy E.; Sivakoff, Gregory R.; Johnson, Kelsey E.; Brogan, Crystal L., 2011, Nature, 470, 66.


\bibitem[\protect\citeauthoryear{Reines et al.}{2013}]{reines13} Reines, Amy E.; Greene, Jenny E.; Geha, Marla, 2013, 775, 116.


\bibitem[\protect\citeauthoryear{Repetto et al.}{2017}]{reppeto} Reppeto, R., Mart\'inez--Garc\'ia, E. E., Rosado, M. \& Gabbasov, R., 2017, MNRAS, 468, 180.

\bibitem[\protect\citeauthoryear{Reunanen, Kotilainen \& Prieto}{2002}]{reunanen02} Reunanen, J., Kotilainen, J. K., \& Prieto, M. A., 2002, MNRAS, 331, 154. 


\bibitem[\protect\citeauthoryear{Reynes et al.}{2009}]{reynes} Rayner, J.T.; Cushing, M.C.; \& Vacca, W.D. 2009, ApJS, 185, 289. 


\bibitem[\protect\citeauthoryear{Riffel et al.}{2006}]{eso428} Riffel R. A., Sorchi-Bergmann T., Winge C., Barbosa F. K. B., 2006a, MNRAS, 373, 2.

\bibitem[\protect\citeauthoryear{Riffel et al.}{2008}]{n4051} Riffel, Rogemar A., Storchi-Bergmann, T., Winge, C., McGregor, P. J., Beck, T., Schmitt, H. 2008, MNRAS, 385, 1129.

\bibitem[\protect\citeauthoryear{Riffel et al.}{2009}]{rr09} Riffel, Rog\'erio; Pastoriza, Miriani G.; Rodr\'iguez-Ardila, A.; Bonatto, C., 2008, MNRAS, 400, 273.

\bibitem[\protect\citeauthoryear{Riffel et al.}{2010a}]{mrk1066-exc} Riffel, R. A.., Storchi-Bergmann, T. \& Nagar, N. M., 2010, MNRAS, 404, 166.


\bibitem[\protect\citeauthoryear{Riffel et al.}{2010b}]{mrk1066-pop} Riffel, Rogemar A.; Storchi-Bergmann, Thaisa; Riffel, Rog\'erio; Pastoriza, Miriani G., 2010, ApJ, 713, 469.


\bibitem[\protect\citeauthoryear{Riffel}{2010c}]{profit} Riffel R. A., 2010c, Ap\&SS, 327, 239.




\bibitem[\protect\citeauthoryear{Riffel, Storchi-Bergmann, \& Winge}{2013}]{mrk79} Riffel, R. A., Storchi-Bergmann, T., \& Winge, C., 2013, MNRAS, 430, 2249

\bibitem[\protect\citeauthoryear{Riffel}{2013}]{baaa} Riffel, R. A. 2013, BAAA, 56, 13. 


\bibitem[\protect\citeauthoryear{Riffel et al.}{2014}]{ngc1068-exc} Riffel, R. A., Vale, T. B., Storchi-Bergmann, T., McGregor, P. J., 2014, MNRAS, 442, 656.



\bibitem[\protect\citeauthoryear{Riffel et al.}{2015a}]{riffel15} Riffel, R. A. et al. 2015, MNRAS, 446, 2823.


\bibitem[\protect\citeauthoryear{Riffel et al.}{2017}]{llp-stel} Riffel, R. A. et al. 2017, MNRAS, 470, 992.

\bibitem[\protect\citeauthoryear{Riffel et al.}{2018}]{llp-sample} Riffel, R. A.  et al., 2018, MNRAS, 474, 1373.





\bibitem[\protect\citeauthoryear{Riffel et al.}{2011}]{rogerio11} Riffel, Rog\'erio; Riffel, Rogemar A.; Ferrari, Fabricio; Storchi-Bergmann, Thaisa, 2011, MNRAS, 416, 493

\bibitem[\protect\citeauthoryear{Riffel et al.}{2013}]{rogerio13} Riffel, R.; Rodr\'iguez-Ardila, A.; Aleman, I.; Brotherton, M. S.; Pastoriza, M. G.; Bonatto, C.; Dors, O. L., 2013, MNRAS, 430, 2002

\bibitem[\protect\citeauthoryear{Riffel et al.}{2015b}]{rogerio15} Riffel, R. et al. 2015, MNRAS, 450, 3069.


\bibitem[\protect\citeauthoryear{Rodr\'iguez-Ardila et al.}{2004}]{ardila04} Rodr\'iguez-Ardila, A.; Pastoriza, M. G.; Viegas, S.; Sigut, T. A. A.; Pradhan, A. K., 2004, A\&A, 425, 457

\bibitem[\protect\citeauthoryear{Rodr\'iguez-Ardila, Riffel \& Pastoriza}{2005}]{ardila05} Rodr\'iguez-Ardila, A.; Riffel, R.; Pastoriza, M. G., 2005, MNRAS, 364, 1041

\bibitem[\protect\citeauthoryear{Rodr\'iguez-Ardila et al.}{2017}]{ardila17} Rodr\'iguez-Ardila, A., Prieto, M. A., Mazzalay, X., Fern\'andez-Ontiveros, J. A., Luque, R., M\"uller-S\'anchez, F., 2017, MNRAS, 470, 2845.
	

\bibitem[\protect\citeauthoryear{Sanchez et al.}{2015}]{sanchez15} Sanchez, S. et. al. 2015, A\&A, 574, A47.

\bibitem[\protect\citeauthoryear{S\'anchez Almeida et al.}{2015}]{sanchez-almeida} S\'anchez Almeida, J., Elmegreen, B. G., Mu\"noz-Tu\"n
\'on, C., Elmegreen, D. M., P\'erez-Montero, E., Amor\'in, R., Filho, M. E., Ascasibar, Y., Papaderos, P., V\'ilchez, J. M. 2015, ApJ, 810, L15. 
 

\bibitem[\protect\citeauthoryear{Sarzi et al.}{2010}]{sarzi10} Sarzi, M., et. al. 2010, MNRAS, 402, 2187.

\bibitem[\protect\citeauthoryear{Scoville et al.}{1982}]{scoville} Scoville, N. Z., Hall, D. N. B., Kleinmann, S. G., \& Ridgway, S. T. 1982, 253, 136.

\bibitem[\protect\citeauthoryear{Sch\"onel et al.}{2014}]{astor14} Sch\"onel, A. J., Riffel, R. A., Stochi-Bergmann, T., Winge, C., 2014, MNRAS, 445, 414.

\bibitem[\protect\citeauthoryear{Sch\"onel et al.}{2017}]{astor17} Sch\"onel, A. J., Riffel, R. A., Stochi-Bergmann, T., Riffel, R., 2017,
MNRAS, 464, 1771.

\bibitem[\protect\citeauthoryear{Schnorr--M{\"u}ller et al.}{2014a}]{allan14} Schnorr M{\"u}ller A., Storchi-Bergmann T., Nagar, N. M., \& Ferrari, F. 2014a, MNRAS, 438, 3322


\bibitem[\protect\citeauthoryear{Schnorr--M{\"u}ller et al.}{2014b}]{allan14b} Schnorr M{\"u}ller A., Storchi-Bergmann T., Nagar, N. M., \& Robinson, A., Lena, D., Riffel, R. A., Couto, G. S., 2014b, MNRAS, 437, 1708.

	
\bibitem[\protect\citeauthoryear{Stasi\'nska et al.}{2015}]{stasinska15} Stasi\'nska, G., Costa-Duarte, M. V., Vale Asari, N., Cid Fernandes, R., Sodr\'e, L., 2015, MNRAS, 449, 559.

\bibitem[\protect\citeauthoryear{Sternberg \& Dalgarno}{1989}]{sternberg89} Sternberg, A. \& Dalgarno, A., 1989, ApJ, 338, 197.

\bibitem[\protect\citeauthoryear{Storchi-Bergmann et al.}{2009}]{stb09} Storchi-Bergmann, T., McGregor, P. J., Riffel, R. A., Sim{\~o}es Lopes, R., Beck, T., Dopita, M., 2009, MNRAS, 394, 1148.
	
\bibitem[\protect\citeauthoryear{Singh et al.}{2013}]{singh13} Singh, R. et al., 2013, A\&A, 558, 43.

\bibitem[\protect\citeauthoryear{Shih \& Fabian}{2003}]{shih} Shih, D. C.; Iwasawa, K.; Fabian, A. C. 2003, MNRAS, 341, 973


\bibitem[\protect\citeauthoryear{Tody}{1986}]{tody86} Tody, D. 1986, The IRAF Data Reduction and Analysis System in Proc. SPIE Instrumentation in Astronomy VI, ed. D.L. Crawford, 627, 733


\bibitem[\protect\citeauthoryear{Tody}{1993}]{tody93} Tody, D. 1993, IRAF in the Nineties in Astronomical Data Analysis Software and Systems II, A.S.P. Conference Ser., Vol 52, eds. R.J. Hanisch, R.J.V. Brissenden, J. Barnes, 173.


\bibitem[\protect\citeauthoryear{Veilleux \& Osterbrock}{1987}]{veilleux} Veilleux, Sylvain and Osterbrock, E. D. 1987, Spectral Classification of Emission-Line Galaxies, ApJ\&SS, 63, 295

\bibitem[\protect\citeauthoryear{Wrobel \& Ho}{2006}]{wrobel} Wrobel, J. M.; Ho, L. C., 2006, ApJL, 646, 95


\end{thebibliography}
\end{document}